\documentclass[pra,reprint]{revtex4-2}
\usepackage{amsthm}
\usepackage{amsmath}
\usepackage{latexsym}
\usepackage{amsfonts}
\usepackage{amssymb}
\usepackage{color}
\usepackage{bm,dsfont}
\usepackage{graphicx}
\usepackage{hyperref}
\usepackage{subfigure}
\usepackage{multirow}
\usepackage{xr}
\usepackage{mathrsfs}
\usepackage[noend]{algpseudocode}
\usepackage{algorithmicx,algorithm}
\usepackage{graphicx}
\usepackage{subfigure}
\usepackage{[number}
\usepackage{float}
\usepackage{svg}

\definecolor{gray}{RGB}{123,123,123}
\newtheorem{theorem}{Theorem}
\newtheorem{proposition}[theorem]{Proposition}%[section]

\theoremstyle{definition}

\newcommand{\hi}{\mathcal{H}} %Hilbert space H
\newcommand{\ket}[1]{|#1\rangle} %ket
\newcommand{\bra}[1]{\langle#1|} %bra
\newcommand{\tr}[1]{\textrm{tr}\left[#1\right]} %trace
\begin{document}
\title{Improved criteria of detecting multipartite entanglement structure}
\author{Kai Wu}
\affiliation{School of Science, Jimei University, Xiamen 361021,China}
\author{Zhihua Chen}
\thanks{chenzhihua77@sina.com}
\affiliation{School of Science, Jimei University, Xiamen 361021,China}
\author{Zhen-Peng Xu}
\thanks{zhen-peng.xu@ahu.edu.cn}
\affiliation{School of Physics and Optoelectronics Engineering,
Anhui University, Hefei 230601, China}
\author{Zhihao Ma}
\thanks{mazhihao@sjtu.edu.cn}
\affiliation{School of Mathematical Sciences, MOE-LSC, Shanghai Jiao Tong University, Shanghai 200240,
China}
\affiliation{Shanghai Seres Information Technology Co., Ltd, Shanghai 200040, China}
\affiliation{Shenzhen Institute for Quantum Science and Engineering, Southern University of Science and Technology, Shenzhen 518055, China}
\author{Shao-Ming Fei}
\thanks{feishm@cnu.edu.cn}
\affiliation{School of Mathematical Sciences, Capital Normal University, Beijing 100048, China}
\affiliation{Max Planck Institute for Mathematics in the Sciences, 04103 Leipzig, Germany}

\begin{abstract}
{Multipartite entanglement is one of the crucial resources in quantum information processing tasks such as quantum metrology, quantum computing and quantum communications. It is essential to verify not only the multipartite entanglement, but also the entanglement structure  in both fundamental theories and the applications of quantum information technologies. However, it is proved to be challenging to detect the entanglement structures, including entanglement depth, entanglement intactness and entanglement stretchability, especially for general states and large-scale quantum systems. By using the partitions of the tensor product space we propose a systematic method to construct powerful entanglement witnesses which identify better the multipartite entanglement structures. Besides, an efficient algorithm using semi-definite programming and a gradient descent algorithm are designed to detect entanglement structure from the inner polytope of the convex set containing all the states with the same entanglement structure. We demonstrate by detailed examples that our criteria perform better than other known ones. Our results may be applied to many quantum information processing tasks.}
\end{abstract}

\maketitle

\section{Introduction}

Multipartite quantum entanglement is crucial in quantum communication and information processing tasks {including quantum computing, quantum cryptography and quantum  teleportation \cite{PhysRep}. Very different from the bipartite case, multipartite quantum entanglement has a rich structure.} Two important concepts have been introduced to characterize such multipartite entanglement structure, i.e., the entanglement producibility and entanglement partitionability \cite{Anders,Otfried}. %{A multipartite mixed entangled quantum state can be decomposed into pure states which are products of entangled states in some subsystems.}
{ Intuitively for pure states,
the partitionability refers to the number of subsystems separable from each other in the multipartite system, while the entanglement producibility, refers to the size of the largest entangled subsystem.  Recently, the entanglement stretchability has been coined \cite{Szalay2019}, which stands for the difference between the entanglement producibility and entanglement partitionability.}

{ Quantum entanglement structure} has many applications, e.g., in spin chains and quantum metrology \cite{Guhne,Otfried,Geza,Zhihong}. However, as the number of subsystems grows, verifying the entanglement structure becomes a hard problem \cite{Anders,Otfried}. Though one can detect theoretically the entanglement structure via state tomography, it becomes infeasible in practice due to the exponential growth of the dimensions of the underlying Hilbert space. { Significant efforts have been devoted in this direction to develop feasible techniques}, such as the utilization of more convenient inequalities given by spin-squeezing and the entries of density matrices, local uncertainty relations, Wigner-Yanase skew information, Fisher information and machine learning\cite{PhysRep,Taming,Huber,Bae,Graph,Jens,Hong_2015,Chen_2018,Hong_2016,Gao_2014,Hong_2021,Vitagliano_2011,Chen_2005,Hyllus_2012,Geza,Gessner,Hong_2021,Chen_2021,Zhou,ChenJiYao_2016,Bancal_2011,Collins_2002,Aloy_2019,LinPeiSheng_2019,LiangYeongCherng_2015,Tura_2019,Lewenstein_2022}.
Gradually, entanglement structure has also been detected experimentally \cite{Christian,Bernd,Hosten2016,McConnell2015,Florian,Zou_2018,Lu, Tian} for special cases, like for Dicke states~\cite{Bernd}. %{\color{blue} However, the current method can not detect the entanglement structure for general states.}

Recently, a systematic approach based on Young diagrams and the partitions of the tensor product space has been proposed to classify the entanglement structures\cite{Szalay2019,Toth2020}. Based on these works,
%we realize that if a smaller set of partitions can be identified, the approach becomes simpler and the resulting criteria is optimized. We need only consider the maximal elements with respect to refinement among all partitions. Therefore,
we propose a systematic method to construct linear criteria to verify the quantum entanglement structures in this paper, %. These witnesses are simply given by local observables and
where the number of measurements needed grows polynomially with the system size in experiments.% {\color{orange} Witnesses verify the entanglement structure from the outer space of the convex set containing the same entanglement structure, providing the necessary conditions.}

{On the other hand, an algorithm based on the adaptive polytope approximation has been proposed to certify quantum separability of bipartite and multipartite quantum systems~\cite{ohst2023certifying}, each iteration is a semidefinite programming (SDP).} {{This algorithm updates each polytope in every iteration to ultimately achieve a satisfactory result, and depends on well-chosen initial polytopes so as to obtain a good enough result. Inspired by this algorithm, we proposed a new algorithm by employing the gradient descent method, which does not depend very much on the initial polytopes in obtaining a good enough result}}.

\section{PRELIMINARY}
{A partition of a set is to classify each element into one and only one subset. For two partitions $\gamma$ and $\xi$, we say $\gamma$ is a refinement of $\xi$, denoted as $\gamma \preceq \xi$, if any subset in $\gamma$ in a subset of the one in $\xi$.
For a given pure state  $\ket{\psi}$  in the form that $\ket{\psi}=\bigotimes_{i} \ket{\psi_{X_i}}$ where $\{X_i\}$ is a partition of the set of all parties, its entanglement structure can be represented by this partition.
If an entanglement structure is representable by a partition $\gamma$ and $\gamma$ is a refinement of $\xi$, then it is also representable by the partition $\xi$.% An example is provided in Fig. \ref{fig:partition}.

Denote $|S|$ the size of the set $S$. For a given pure state  $\ket{\psi}=\bigotimes_{i} \ket{\psi_{X_i}}$ and $\max |X_i| \le k$, where $\{X_i\}$ is a partition of the set of all parties, $\ket{\psi}$  is said to be
\begin{enumerate}
    \item $k$-producible, if  $\max |X_i|\le k$;
    \item $k$-partitionable, if $|\{X_i\}| \le k$;
    \item $k$-stretchable, if $\max |X_i| - |\{X_i\}| \le k$.
\end{enumerate}}
%Similarly, $\ket{\psi}$  is said to be $k$-partitionable if the partition $\{X_i\}$ is formed by no more than $k$ subsets, i.e., $|\{X_i\}| \le k$. In the case that $\max |X_i| - |\{X_i\}| \le k$, $\ket{\psi}$  is said to be $k$-stretchable.
%A partition of a set is a grouping of its elements into non-empty subsets, in such a way that every element is included in exactly one subset. A partition $\gamma$ is a refinement of a partition $\xi$ if every part of $\gamma$ is a subset of some part of $\xi$, denoted as $\gamma \preceq \xi$.

%Consider a pure quantum state $\ket{\psi}=\bigotimes_{i} \ket{\psi_{X_i}}$, where $\{ X_i \}$ represents a set of subsystems. It is evident that the collection of subsystems $\{ X_i \}$ constitutes a partition of the composite system.
%$\ket{\psi}$ is termed $k$-producible if each component state $\ket{\psi_{X_i}}$ is a state {\color{orange} of} at most $k$ partites. Moreover, $\ket{\psi}$ is referred to as $k$-partitionable if the partition $\{ X_i \}$ contains at least $k$ parts. Additionally, $\ket{\psi}$ is said to be $k$-stretchable if the cardinality of the largest subset $X_i$ does not exceed the number of parts in the partition $\{ X_i \}$ by more than $k$.
{ A mixed state $\rho$ is called $k$-producible if it can be decomposed as a convex combination of $k$-producible pure states. Similarly, the definition of $k$-partitionable, or $k$-stretchable can be extended to mixed states.}

\begin{figure}
\includegraphics[width=1.0\columnwidth]{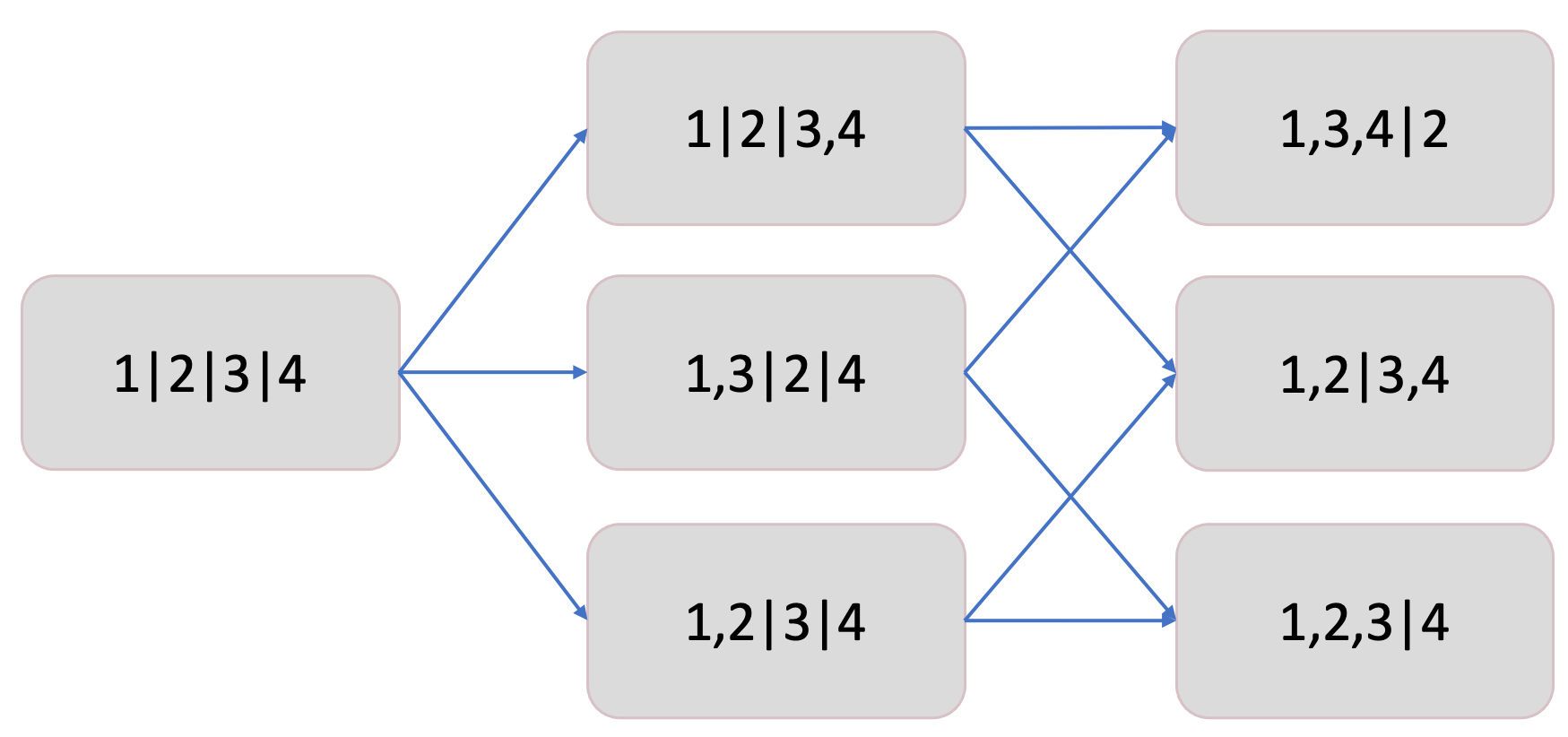}
\caption{\label{fig:partition}  Seven partitions of $\{1,2,3,4\}$ are $\gamma_1=1|2|3|4$, $\gamma_2=1|2|3,4$, $\gamma_3=1,3|2|4$, $\gamma_4=1,2|3|4$, $\gamma_5=1,3,4|2$, $\gamma_6=1,2|3,4$ and $\gamma_7=1,2,3|4$. The arrows in the diagram denote the refinement relationship, denoted by $\preceq$. For the partition set $\Gamma$ consists of all seven partitions, $\Gamma^{\max} = \{ \gamma_5, \gamma_6, \gamma_7 \}$.}
\end{figure}

{Given a collection $\Gamma$ of partitions, the set of maximal elements under the partial order of refinement $\preceq$ is denoted $\Gamma^{\max}$. 
To continue, we introduce further notations:
\begin{align}
    &\Gamma_{k-\text{part}} = \{\xi = \{X_i\} : |\{X_i\}| \le k\},\\
    &\Gamma_{k-\text{prod}} = \{\xi = \{X_i\} : \max |X_i| \le k\},\\
    &\Gamma_{k-\text{str}} = \{\xi = \{X_i\} : \max |X_i| - |\{X_i\}| \le k\}.
\end{align}
Consequently, $\Gamma_{k-\text{prod}}^{\max}$, $\Gamma_{k-\text{part}}^{\max}$ and $\Gamma_{k-\text{str}}^{\max}$ can be defined.
If the entanglement structure of a pure quantum state is representable by a partition within $\Gamma$, it is necessarily also representable by a partition within $\Gamma^{\max}$. An example is depicted in Fig. \ref{fig:partition} with $\Gamma$ to be the set of all the partitions of $\{1,2,3,4\}$.}

\section{Detection of entanglement structure}

Consider a direct product quantum state pairs $(\ket{m}, \ket{n})$ with $\ket{m} = \bigotimes_{i=1}^N \ket{m_i}$, $\ket{n} = \bigotimes_{i=1}^N \ket{n_i}$, where $\ket{m_i}$ and $\ket{n_i}$ are the standard orthogonal basis of $\hi_i$. We have the following result,{ which pertains to an inequality involving a off-diagonal element of a density matrix and two of its diagonal elements:
\begin{proposition} \label{prop:1}
For any pure state $\ket{\psi}=\bigotimes_{X_i \in \gamma} \ket{\psi_{X_i}}$ and $\forall X_i\in\gamma$, we have
\begin{align}\label{neq:one-simple}
\begin{aligned}
|\bra{m} \psi\rangle \bra{\psi} n\rangle| \leq & \frac{1}{2} [ (\bra{m_{\bar{X}_i}} \bra{n_{X_i}}) \ket{\psi} \bra{\psi} (\ket{m_{\bar{X}_i}} \ket{n_{X_i}})  + \\
&(\bra{m_{X_i}} \bra{n_{\bar{X}_i}}) \ket{\psi} \bra{\psi} (\ket{m_{X_i}} \ket{n_{\bar{X}_i}}) ],
\end{aligned}
\end{align}
where $\bar{X}_i$ stands for the complement set of the set $X_i$.
\end{proposition}}
The proof is given in Appendix I. Violation of the inequality in Eq.~\eqref{neq:one-simple} means that $\ket{\psi} \neq \bigotimes_{X_i \in \gamma} \ket{\psi_{X_i}}$.

{Based on the inequality in Eq.~\eqref{neq:one-simple}, we continue to construct further criteria involving more off-diagonal elements.}
\subsection{Criteria based on the outer polytopes}
{Firstly, we take a $3$-qubit pure quantum state $\rho = \ket{\psi} \bra{\psi}$ to illustrate the whole procedure.
Consider the set of off-diagonal elements $P = \{\rho_{2,3}, \rho_{2,5}, \rho_{3,5} \}$ and the partition $12|3$. If $\ket{\psi}=\ket{\psi_{12}} \otimes \ket{\psi_3}$, a sequence of inequalities can be derived from Proposition \ref{prop:1} as follows,
\begin{eqnarray}\label{eq:partition:12:3}
\begin{cases}
|\rho_{2, 3}| - \frac{1}{2}  \rho_{1,1} - \frac{1}{2}  \rho_{4,4} \le 0, \\
|\rho_{2, 5}| - \frac{1}{2}  \rho_{1,1} - \frac{1}{2}  \rho_{6,6} \le 0, \\
|\rho_{3, 5}| - \frac{1}{2}  \rho_{3,3} - \frac{1}{2}  \rho_{5,5} \le 0.
\end{cases}
\end{eqnarray}
Besides, any diagonal element of the state $\rho$ is non-negative and the sum of the appeared diagonal elements is no more than $1$.
These inequalities define a series of closed half-spaces, whose intersection results in a convex polytope \cite{grunbaum1967convex}, denoted by $\Theta_{12|3}(P)$. Thus, if $\ket{\psi}$ can be decomposed as the tensor product $\ket{\psi_{12}} \otimes \ket{\psi_3}$, we must have the corresponding elements of the density operator $\rho = |\psi\rangle\langle \psi|$ to be in the polytope $\Theta_{12|3}(P)$. Consequently, any mixed state of the same entanglement structure is also contained in $\Theta_{12|3}(P)$.

Similarly, we could construct another two polytopes $\Theta_{1|23}(P)$ and $\Theta_{2|13}(P)$ for the partitions $1|23$ and $2|13$, respectively. Denote $\Theta_{2-\text{prod}}(P)$ the convex hull of all the points in $\Theta_{12|3}(P)$, $\Theta_{1|23}(P)$ and $\Theta_{2|13}(P)$. Any mixed $2$-producible states must be in $\Theta_{2-\text{prod}}(P)$.

In general, given a partition set $\Gamma$ and a collection $P$ of non-diagonal elements from a density matrix, Proposition \ref{prop:1} can be employed to derive inequalities that define a convex polytope $\Theta_{\gamma}(P)$ for each maximum partition $\gamma$ within $\Gamma$, i.e., each partition in $\Gamma^{\max}$. The smallest convex polytope that encompasses all $\Theta_{\gamma}(P)$ is referred to as $\Theta_{\Gamma}(P)$.
The set $\Gamma$ can be either $\Gamma_{k-\text{prod}}$, $\Gamma_{k-\text{part}}$, or $\Gamma_{k-\text{str}}$, each capturing a distinct aspect of the entanglement structure. The explicit result is summarized as follows.}

{
\begin{theorem} \label{theorem:1}
Let $\rho = \sum_{i} p_i \ket{\psi_i}\bra{\psi_i}$ be an $N$-partite state  with $\sum_i p_i = 1$ and $p_i> 0$. If there exists $\xi_i \in \Gamma$ such that $\ket{\psi_i} = \bigotimes_{X_j \in \xi_i} \ket{\psi_{i,{X_j}}}$ for all $\ket{\psi_i}$, we have
\begin{equation}
    \rho \in \Theta_{\Gamma}(P),
\end{equation}
where $\Theta_{\Gamma}(P)$ is the polytope constructed from a subset of all possible inequalities that can be formulated from $\Gamma^{\max}$.
\end{theorem}
}
{In practice, one can employ} the Polyhedra.jl package \cite{legat2023polyhedral} in Julia to {carry out operations on polytope such as intersection and  convex hull of several polytopes. An example with the mixed $4$-qubit W-state $\rho_{W_4}$ is presented in Section III.D.}

\subsection{Criteria based on the hyperplanes}

Due to the fact that the computational complexity of the polytope-based criteria increases fastly as the dimension grows, we propose a simplified criterion based on hyperplanes, { which offers a more friendly computational approach}.

{ Similar as in Subsection A, we still consider the set of non-diagonal elements $P = \{\rho_{2,3}, \rho_{2,5}, \rho_{3,5} \}$ of a $3$-qubit pure quantum state $\rho = \ket{\psi} \bra{\psi}$, and the partitions in $\{12|3, 1|23, 2|13\}$. By simply summing up all the inequalities within $\Theta_{12|3}(P)$ in Eq.~\eqref{eq:partition:12:3}, a unified inequality is created. Similar inequalities can be obtained by applying this method to $\Theta_{2|13}(P)$ and $\Theta_{1|23}(P)$. Those three new inequalities are as follows:
\begin{eqnarray}
  \begin{cases}\label{eq:sumup}
\sum\limits_{\rho_{m,n} \in P} |\rho_{m,n}| \leq \frac{1}{2}(2\rho_{1,1}+\rho_{6,6}+\rho_{7,7}+\rho_{2,2}+\rho_{3,3}), \\
\sum\limits_{\rho_{m,n} \in P} |\rho_{m,n}| \leq \frac{1}{2}(2\rho_{1,1}+\rho_{4,4}+\rho_{7,7}+\rho_{2,2}+\rho_{5,5}), \\
\sum\limits_{\rho_{m,n} \in P} |\rho_{m,n}| \leq \frac{1}{2}(2\rho_{1,1}+\rho_{4,4}+\rho_{6,6}+\rho_{3,3}+\rho_{5,5}).
\end{cases}
\end{eqnarray}
The right-hand side of the set of inequalities can be formulated as $M \cdot \vec{\beta}^T$, where the row vector $\vec{\beta}$ contains all diagonal elements of $\rho$, and $M$ is a coefficient matrix. More explicitly,
\begin{equation*}
    \vec{\beta} = [ \rho_{1,1}, \rho_{2,2}, \rho_{3,3}, \rho_{4,4}, \rho_{5,5}, \rho_{6,6}, \rho_{7,7}, \rho_{8,8} ],
\end{equation*}
\begin{equation*}
    M = \begin{bmatrix}
       1 & 1/2 & 1/2 & 0 & 0 & 1/2 & 1/2 & 0\\
       1 & 1/2 & 0 & 1/2 & 1/2 & 0 & 1/2 & 0 \\
       1 & 0 & 1/2 & 1/2 & 1/2 & 1/2 & 0 & 0 \\
    \end{bmatrix}.
\end{equation*}

By applying a column-wise maximum operation to matrix $M$, we obtain a vector $\vec{\alpha}$:
\begin{equation*}
    \vec{\alpha} = \begin{bmatrix}
    1 & 1/2 & 1/2 & 1/2 & 1/2 & 1/2 & 1/2 & 0
    \end{bmatrix}.
\end{equation*}
Consequently, we have a new inequality
\begin{equation}
  \sum\limits_{\rho_{m,n}\in P} |\rho_{m,n}| \le \vec{\alpha}.\vec{\beta}^T,
\end{equation}
which is a relaxation of all the three inequalities in Eq.~\eqref{eq:sumup}.
More explicitly, the new derived criterion is
\begin{align}
  & |\rho_{2,3}|+|\rho_{2,5}|+|\rho_{3,5}| \nonumber \\
  \leq\ & \rho_{1,1} + ( \rho_{2,2}+\rho_{3,3}+\rho_{4,4}+\rho_{5,5}+\rho_{6,6}+\rho_{7,7})/2.
\end{align}

In general, for a given set of non-diagonal elements denoted as $ P $ and the partition set $\Gamma$, we can only consider all the partitions in $\Gamma^{\max}$ and carry out the same procedure to obtain a new criterion. The main steps are summarized as follows.

\begin{enumerate}
  \item For each partition $\gamma \in \Gamma^{\max}$, and $\rho_{m,n}\in P$, we obtain from Proposition 1 the inequalities that
    \begin{equation}\label{eq:step1}
    |\rho_{m,n}| \le (\rho_{d_{\gamma,1},d_{\gamma,1}} + \rho_{d_{\gamma,2},d_{\gamma,2}})/2,
    \end{equation}
    where $d_{\gamma,1}$ and $d_{\gamma,2}$ depend on $m,n$ and the partition $\gamma$. We remark that there could be different inequalities for the same pair of $(m,n)$.
  \item By choosing one inequality as in Eq.~\eqref{eq:step1} for each $\rho_{m,n}\in P$ and a given partition $\gamma$, we obtain
  \begin{equation}
    \sum_{\rho_{m,n}\in P} |\rho_{m,n}| \le \vec{m}_\gamma.\vec{\beta}^T,
  \end{equation}
  where $\vec{m}_\gamma$ corresponds to a row in the matrix $M$.
\item By applying a column-wise maximum operation to matrix $M$, we obtain a vector $\vec{\alpha}$:
  \begin{equation}
  \label{maininequ}
    \sum_{\rho_{m,n}\in P} |\rho_{m,n}| \le \vec{\alpha}.\vec{\beta}^T,
  \end{equation}
\end{enumerate}
The partition set $\Gamma$ can correspond to either $\Gamma_{k-\text{prod}}$, $\Gamma_{k-\text{part}}$ or $\Gamma_{k-\text{str}}$.

Since there could be different options at the second step, the goal to construct the vector $\vec{\alpha}$ is to minimize the sum of all elements in vector $\vec{\alpha}$. To achieve this goal, we can follow the principle to have as many common non-zero columns as possible for each row in the matrix during the construction of $M$, i.e., the choice of inequality for each pair $(m,n)$. In other words, we aim to reduce the number of appeared non-zero columns. This strategy helps us to effectively utilize the structure of matrix $M$ while keeping the sum of the elements in vector a minimal. This approach is beneficial for our optimization process.

Further elaboration on the optimization process is provided in Appendix II. The greedy algorithm is implemented to obtain the optimal inequalities, see the codes listed in \cite{Wu}.}

{The left hand side of the inequality
(\ref{maininequ}) can be replaced by its lower bound $\sum\limits_{\rho_{m,n}\in P}\frac{1}{2}(\rho_{m,n}+\rho_{n,m})$, which can be reresented by the expectation value of the observable $W_{P}=\frac{1}{2}\ket{m}\bra{n}+\ket{m}\bra{n}$. On the other hand, the right hand side of (\ref{maininequ}) can be represented the expectation value of the observable $W=diag\{\vec{\alpha\}}$}. Our hyperplane-based criteria is experimentally friendly, which can be implemented by measuring the two observables without state tomography.  The two observables can be realized by using local observables and the number of the required local observables scales polynomially with the system size. In comparison, to realize full state tomography, the number of the required local observables grows exponentially.

\subsection{Criteria based on the inner polytopes}

{An SDP-based method was also developed to create evolving polytopes, whose nodes are different kinds of separable states, for confirming quantum entanglement~\cite{ohst2023certifying}.}
{Besides, a modified version of Gilbert's algorithm was recently used to solve the convex membership problem \cite{PhysRevLett.120.050506, PhysRevA.96.032312}}.
{ The performance of this SDP-based algorithm depends on the choice of the initial polytope, and the current algorithm generates the initial polytope randomly. As the complexity increases with the increasing system, inspired by the previous works\cite{ohst2023certifying,PhysRevLett.120.050506,PhysRevA.96.032312}, we propose a systematic approach in the following to select the initial polytope so as to guarantee the efficiency and performance.}

\begin{figure}
\includegraphics[width=1.0\columnwidth]{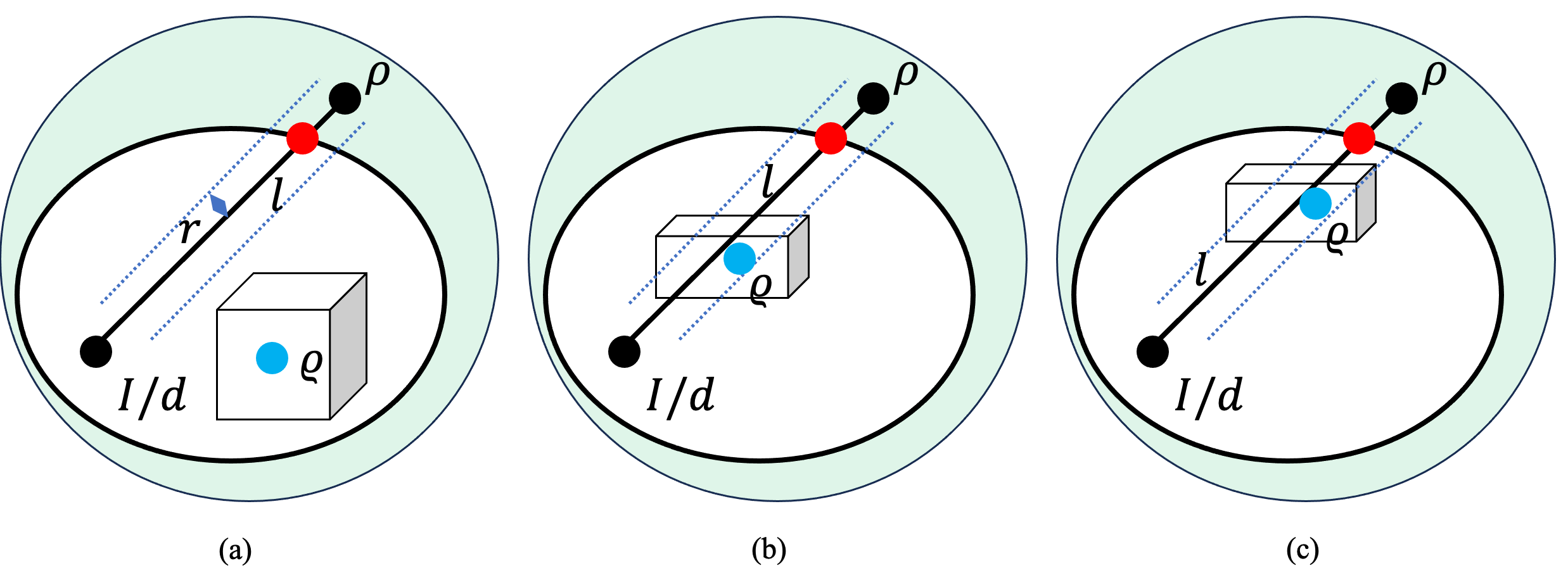}
\caption{\label{fig:next-algo} (a) Initialise an arbitrary polytope $\mathcal{P} = \{ \varrho_i \}$ and $\varrho = \sum_i P_i \varrho_i$. (b) Minimize the distance between the quantum state $\varrho$ and the segment $l$. (c) Minimize the distance between the quantum state $\varrho$ and $\rho$.}
\end{figure}

{ We introduce a new algorithm to certify the entanglement structure of the mixed state {$\rho(t) = t \rho + (1 - t) \mathbb{I}_d / d$} for given $\rho$. Especially, we are interested in the minimal $t$ such that $\rho(t)$ still have the targeted entanglement structure. Our algorithm involves  two stages as illustrated in Fig.~\ref{fig:next-algo}.} {Different from \cite{ohst2023certifying}, initially, we employ gradient descent at the first stage},
%to geometrically converge the polytope $\mathcal{P}$—characterized by its $k$-producible ($k$-partitionable, $k$-stretchable) vertices—towards the state $\rho$.
which is followed by SDP at the second stage. %to optimize the approximation and determine the optimal mixing parameter $t_0$.
The gradient descent is conducted within the Pytorch framework, while the SDP is facilitated through the PICOS \cite{PICOS} library. The whole procedure can be summarized as follows.

\begin{enumerate}
\item Initialize an arbitrary polytope $\mathcal{P}$ determined by a set of states $\{ \varrho_i \}$ of the entanglement structure corresponding to a partition $\gamma$, and a set of probabilities $\{ p_i \}$, where $\sum_i p_i = 1$. 
An example is illustrated in Fig. \ref{fig:next-algo}(a).

\item Employing the gradient descent algorithm, we minimize the geometric distance from the state $\varrho = \sum_i p_i \varrho_i$ to the line segment $l$ between $\rho$ and the maximally mixed state, as shown in Fig. \ref{fig:next-algo}(b).
  
A predefined threshold $r$ serves as the criterion for progression to the next step, ensuring a systematic approach to state optimization.

{\item Using gradient descent to minimize the geometric distance between the state $\varrho$ and $\rho$, ensuring that the polytope's boundaries progressively converge to $\rho$, as illustrated in Fig. \ref{fig:next-algo}(c). The optimization process is directed by a loss function $\lVert \rho - \varrho \rVert = \sqrt{\tr{(\rho - \varrho )^{\dag} (\rho - \varrho)}}$, with the initial polytope configuration determined by the results of step 2. If the distance, as defined in step 2, exceeds the predetermined threshold $r$, the algorithm reverts to step 2 to ensure that the distance between $\varrho$ and the line segment $l$ remains within the specified limit.}
\item Finally, we use SDP to approximate the optimal $t_0$. The initial points are chosen as the results in step 3, but we choose one subsystem from each vertex of $\mathcal{P}$ as the unknown one, which is optimized using SDP. Other subsystems except this one from each vertex are denoted as a new set $\mathcal{P}^{\prime}$. We do this for each subsystem iteratively. The details of the SDP are as follows:
\begin{align} \label{SDP}
\begin{aligned}
\max &\ \ t \\
\mathrm{w.r.t.} &\ \ t, \tau_i \succcurlyeq 0 \\
\mathrm{s.t.} &\ \ \rho(t) = \sum_{\varrho_i \in \mathcal{P}^{\prime}} \varrho_i \otimes \tau_i,
\end{aligned}
\end{align}
where the trace of $\tau_i$ is not necessarily one.
\end{enumerate}

% In step 2, although we allow $\varrho$ to approach the line segment $l$, we can not guarantee that the polytope $\mathcal{P} \cap l \neq \emptyset$. Therefore the feasibility of SDP can not be guaranteed. For the quantum entanglement structures except full separability, we can optimize the quantum subsystem of each vertex corresponding to the largest part in this partition through SDP to guarantee the feasibility of SDP. For full separability, as well as certain quantum entanglement structures where SDP is infeasible, firstly we can construct a polytope with a few of vertices by utilizing mutually-unbiased bases (MUBs) for each part in the partition, which covers a large  volume, and contains the maximally mixed state. Then we  append these vertices to the set $\mathcal{P}^{\prime}$ to guarantee the feasibility of SDP. We explain the procedure in detail by using full separability as an example.

{While step 2 permits the convergence of $\varrho$ towards the line segment $l$, the intersection of the polytope $\mathcal{P}$ with $l$, and consequently the feasibility of SDP, cannot be assured. To improve on this point, we optimize the quantum subsystems corresponding to the largest partition of each node via SDP at each iteration. For fully separable states and certain entanglement structures where SDP is infeasible, we can also construct an initial polytope using only states derived from mutually-unbiased bases (MUBs), which not only covers a substantial volume but also encapsulate the maximally mixed state. %These vertices are incorporated into the set $\mathcal{P}^{\prime}$ to facilitate the feasibility of SDP, a methodology we illustrate with the example of full separability.
%Let the polytope $\mathcal{P}_1$ be described by a set of $6$ vertices $\{ (\mathbb{I}_2 \pm \sigma_x)/2, (\mathbb{I}_2 \pm \sigma_y)/2, (\mathbb{I}_2 \pm \sigma_z)/2 \}$, where $\sigma_x, \sigma_y, \sigma_z$ are Pauli matrices. Actually, these vertices are the mutually unbiased bases. The polytope $\mathcal{P}_1$ contains $\frac{\mathbb{I}_2}{2}$ in Bloch sphere. Then let the polytope $\mathcal{P}_{N-1}$ be described by a set of vertices as follows:
%$\mathcal{P}_{N-1}$ can be described by $6^{N-1}$ vertices.
More explicitly for the case that the local dimension is $2$, the constraint condition in Eq.~\eqref{SDP} transforms into the following expression:}
\begin{equation*}
    \rho(t) = \sum_{\varrho_i \in (\mathcal{P}^{\prime} \bigcup \mathcal{P}_{N-1})} \varrho_i \otimes \tau_i,
\end{equation*}
where ${\cal P}_{N-1}$ is defined as follows:
\begin{equation}
  \big\{ \bigotimes_{i=1}^{N-1} \rho_i, \rho_i \in \big\{ \frac{(\mathbb{I}_2 \pm \sigma_x)}{2}, \frac{(\mathbb{I}_2 \pm \sigma_y)}{2}, \frac{(\mathbb{I}_2 \pm \sigma_z)}{2} \big\} \big\}.
\end{equation}
Since $\mathcal{P}_{N-1}$ contains the maximally mixed state, the feasibility of SDP can be guaranteed.

\subsection{{Applications of method with inner polytopes}}
{ Firstly,} we study the full separability of the states $\rho(t)=t\rho+(1-t)\mathbb{I}_d / d$, where $\rho$ is any state up to six qubits. The results are computed by the algorithm based on the inner polytope. We compare the results with the known ones in the literature in Table \ref{tab:full}.

\begin{table}[h]
\caption{\label{tab:full} Comparison of the full separability computed by our method and the ones by adaptive polytopes \cite{ohst2023certifying}. The ``-'' in the table signifies that \cite{ohst2023certifying} either did not provide the results or the code provided by \cite{ohst2023certifying} was unable to perform the calculations.
}
\begin{ruledtabular}
\begin{tabular}{lcc}
\textrm{Quantum States}&
\textrm{Our result}&
\textrm{Lower bound of \cite{ohst2023certifying}}\\
\colrule
GHZ(6 Qubits) & 0.030303 & -\\
W(6 Qubits) & 0.02345 & -\\
Cluster state (6 Qubits) & 0.030303 & -\\
GHZ(5 Qubits) & 0.05882 & 0.05878 \\
W(5 Qubits) & 0.047084 & 0.046956 \\
Cluster state (5 Qubits) & 0.05882 & - \\
GHZ(4 Qubits) & 0.1111 & 0.1111\\
W(4 Qubits) & 0.0926 & 0.0926\\
Cluster state (4 Qubits) & 0.1111 & 0.1111\\
Dicke (4 Qubits, 2 ex.) & 0.085708 & 0.08571\\
\end{tabular}
\end{ruledtabular}
\end{table}

For the $4$ qubit, $5$-qubit and $6$-qubit noisy GHZ states, the results computed by our algorithm based on the inner polytope give the necessary and sufficient values of the full separability, which are $1/9$, $1/17$ and $1/33$, respectively \cite{Chen_2018}.
Our method performs better compared with the algorithm in \cite{ohst2023certifying} in most cases.

{Then,} we utilize our algorithm based on the inner polytope to compute the entanglement parititionability and the entanglement producibility of $\rho(t)$ with $\rho$ to be  4-qubit W state, 4-qubit and 5-qubit GHZ states in Tables \ref{tab:w-4-inside}, \ref{tab:ghz-4-inside} and \ref{tab:ghz-5-inside} as follows.

\begin{table}[h]
\caption{\label{tab:w-4-inside} Comparison of the partitionability  and the producibility computed by our algorithm and the known ones in the literature for the mixture of 4-qubit W state and white noise.
}
\begin{ruledtabular}
\begin{tabular}{ccc}
\textrm{Entanglement Structure}&
\textrm{Our result}&
\textrm{Other known results}\\
\colrule
3-part & 0.247 & 0.243\cite{PhysRevLett.120.050506} \\
2-prod & 0.247 & 0.245\cite{PhysRevLett.120.050506} \\
2-part & 0.4705 & 0.474\cite{Taming} \\
\end{tabular}
\end{ruledtabular}
\end{table}

\begin{table}[h]
\caption{\label{tab:ghz-4-inside} Comparison of the partitionability  and the producibility computed by our algorithm and the known ones in the literature for the mixture of 4-qubit GHZ state and white noise.
}
\begin{ruledtabular}
\begin{tabular}{ccc}
\textrm{Entanglement Structure}&
\textrm{Our result}&
\textrm{Other known results}\\
\colrule
3-part & 0.19997 & 0.198\cite{PhysRevLett.120.050506} \\
2-prod & 0.27251 & 0.269\cite{PhysRevLett.120.050506} \\
2-part & 0.46456 & 0.467\cite{Taming}(exact) \\
\end{tabular}
\end{ruledtabular}
\end{table}

{ Table \ref{tab:w-4-inside} shows that our results based on inner polytope perform better than the results in \cite{PhysRevLett.120.050506} in most cases for $4$-qubit noisy W state. Table \ref{tab:ghz-4-inside} shows the bound corresponding to $3$-partitionablility is $0.19997$ for $4$-qubit noisy GHZ state, which approaches the necessary and sufficient value $p=0.2$ \cite{Chen_2018}. Besides, the bound $0.2725$ for $2$-produciblity is better than the one in \cite{PhysRevLett.120.050506}}.

\begin{table}[h]
\caption{\label{tab:ghz-5-inside} Comparison of the partitionability  and the producibility computed by our algorithm and the known ones in the literature for the mixture of 5-qubit GHZ state and white noise.
}
\begin{ruledtabular}
\begin{tabular}{ccc}
\textrm{Entanglement Structure}&
\textrm{Our result}&
\textrm{Other known Results}\\
\colrule
4-part & 0.09434 & 0.0943 \cite{Chen_2018}\\
2-prod & 0.2375  & 0.19\cite{PhysRevLett.120.050506} \\
3-part & 0.2377  & 0.2380\cite{Chen_2018}\\
3-prod & 0.3846 & 0.278\cite{PhysRevLett.120.050506} \\
2-part & 0.48378 & 0.4839\cite{Chen_2018}\\
\end{tabular}
\end{ruledtabular}
\end{table}

{Table \ref{tab:ghz-5-inside} is for $5$-qubit noisy GHZ state. Our results of $2/3/4$-partitionablity are very close to the optimal known values given in \cite{Chen_2018}. For the producibility, our results are better compared with the ones in \cite{PhysRevLett.120.050506}. From \cite{Hong_2021} the exact bound is no more than $0.238$ for $2$-producibility of $\rho(t)$, which is close to our bound $0.2375$.}

{Besides, our algorithm can provide the exact decomposition of $k$-partitionable (producible, or stretchable) states. By calculation the fidelity between the state $\rho(t)$ and the decomposition of the state obtained by our algorithm is $99.999\%$. Thus the precision of our algorithm is guaranteed.
}

{{Furthermore, we can detect the entanglement structure of any state $\rho$ mixed with any fully separable states. For example, we consider $\rho(t)=t \ket{W}\bra{W}+(1-t)\rho_{f}$, where $\ket{W}=(\ket{0001}+\ket{0010}+\ket{0100}+\ket{1000})/2$ and $\rho_{f}=\bigotimes\limits_{i=1}^4\tau_i$ with $\tau_i=(3|0\rangle\langle 0|+|1\rangle\langle 1|)/4$. Our algorithm shows that $\rho(t)$ is fully separable when $t<0.1468$ and 2-partitionable if $t<0.58$, while $\rho(t)$ is fully separable when $t<0.1464$ and 2-partitionable for $t<0.5623$ in \cite{PhysRevLett.120.050506}.}}

{To guarantee the universality of our algorithm, in our algorithm the initial states are optimized by the gradient conjugate method. We only optimize one partite states, given that the other partite states are initialized. In this way, the runtime of our algorithm could be slightly longer than that in \cite{ohst2023certifying}.}

\subsection{Applications of method with hyperplane}

To illustrate our results by the witness based on the hyperplane, we give the following examples.

{\bf{Example 1.}} Consider the family of $4$-qubit quantum states,
\begin{equation*}
    \rho_{W_4} = (1-t) \frac{\mathbb{I}}{2^4} + t \ket{W_4} \bra{W_4},
\end{equation*}
with $\ket{W_4} = \frac{1}{2}(\ket{0001} + \ket{0010} + \ket{0100} + \ket{1000})$.

We select $P$ as the one given in Appendix III and obtain that if $\rho$ is $2$-producible or $3$-partitionable, $0$-stretchable or $(-1)$-stretchable, the entries of $\rho$ fulfill the following inequality,
\begin{equation} \label{eq:W_4_2}
\frac{1}{2} | \textrm{Tr} [W_P \cdot \rho] | \le \textrm{Tr}[W_{2-\text{prod}} \cdot \rho],
\end{equation}
where { $W_P$ select the targeted off-diagonal elements of $\rho$}, and  $W_{2-\text{prod}}=\textrm{diag}\{a_1,\cdots\}$ is the diagonal matrix with $a_1 = 2$ and $a_{2} = a_{3} = a_{4} = a_{5} = a_{6} = a_{7} = a_{9} = a_{10} = a_{11} = a_{13} = 1/2$.

\begin{figure}
\includegraphics[width=1.0\columnwidth]{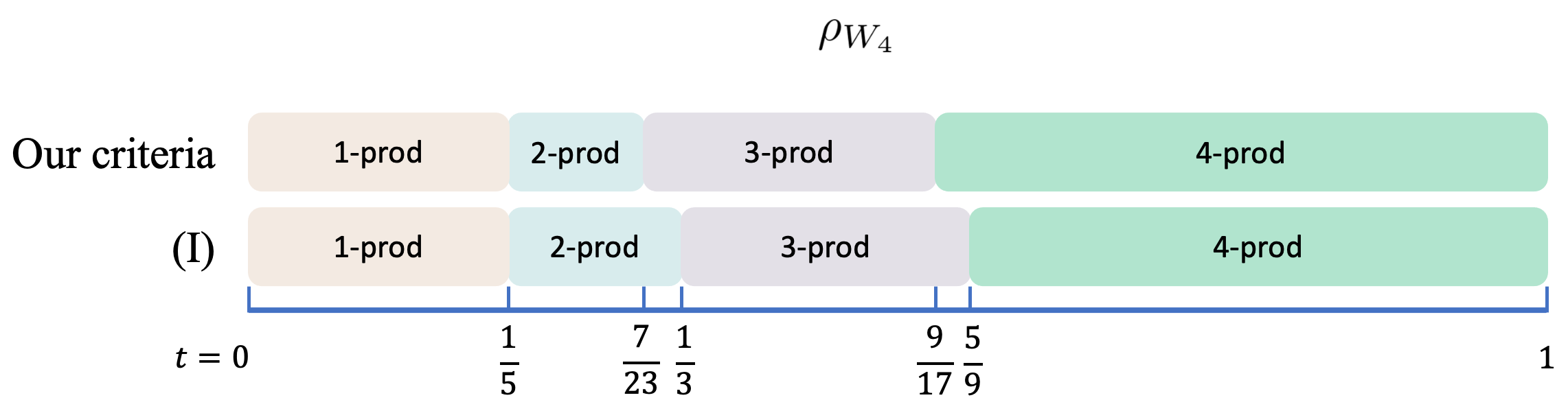}
\caption{\label{fig:w-4} The illustration shows the results obtained by using our criteria and Theorem 2 of Ref. \cite{Hong_2021}((I) in the figure). to detect the entanglement depth.}
\end{figure}

If $\rho$ is $3$-producible or $2$-partitionable, the entries of $\rho$ satisfy
\begin{equation} \label{eq:W_4_3}
\frac{1}{2} | \textrm{Tr} [W_P \cdot \rho] | \le \textrm{Tr}[W_{3-\text{prod}} \cdot \rho],
\end{equation}
where $W_{3-\text{prod}}=\textrm{diag}\{a_1,\cdots\}$ is the diagonal matrix with $a_1 = 2$, $a_{2} = a_{3} = a_{5} = a_{9} = 1$ and $a_{4} = a_{6} = a_{7} = a_{10} = a_{11} = a_{13} = 1/2$. The witness in \cite{Lu} can detect $2$ ($3$ and $4$)-partite entanglement when $p>0.592$ ($p>0.729$ and $p>0.873$)  while our theorem can detect  $2$ ($3$ and $4$)-partite entanglement when $p>0.2$ ($p>0.304$ and $p>0.529$). The results compared with the best result so far in \cite{Hong_2021} are illustrated in Fig. \ref{fig:w-4}. %The results obtained through Theorem \ref{theorem:1} coinside with those obtained through Theorem \ref{theorem:2}.
The details can be found in Appendix III.

\begin{figure}
\includegraphics[width=1.0\columnwidth]{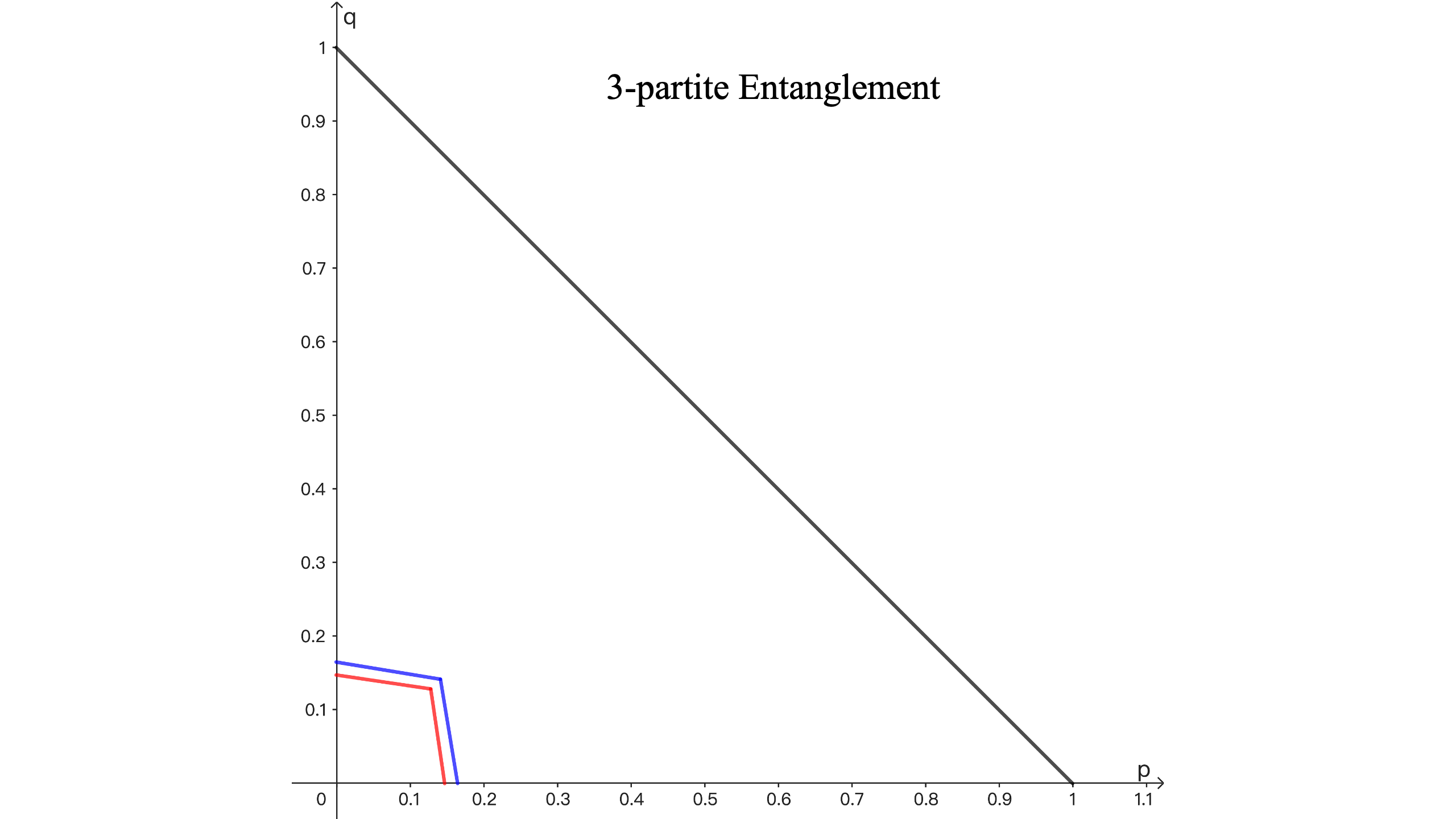}
\caption{\label{fig:compare} The red line consists of the line $q = \frac{14}{95}(1-p)$, $p \in [0, \frac{14}{109}]$ and the line $p = \frac{14}{95}(1-q)$, $q \in [0, \frac{14}{109}]$. The area enclosed by the red line, $p$ axis, line $q = 1 - p$ and the $q$ axis corresponds to the quantum states containing genuine $3$-partite entanglement detected by our Theorem II. The blue line consists of the line $q = \frac{16}{97}(1-p)$, $p \in [0, \frac{16}{113}]$ and the line $p = \frac{16}{97}(1-q)$, $q \in [0, \frac{16}{113}]$. The area enclosed by the blue line, $p$ axis, line $q = 1 - p$ and the $q$ axis corresponds to the quantum states containing genuine $3$-partite entanglement detected by the Theorem 2 of Ref.\cite{Hong_2021}.}
\end{figure}

{\bf{Example 2.}} Now we consider the family of 4-qutrit quantum states,
\begin{equation*}
    \rho(p, q) = p \ket{W}\bra{W} + q \sigma^{\otimes 4} \ket{W} \bra{W} \sigma^{\otimes 4} + \frac{1-p-q}{3^4} \mathbb{I},
\end{equation*}
where $\ket{W} = \frac{1}{2\sqrt{2}}(\sum\limits_{i=1}^{2} \ket{i000} + \ket{0i00} + \ket{00i0} + \ket{000i})$, and $\sigma \ket{0} = \ket{1}$, $\sigma \ket{1} = \ket{2}$, $\sigma \ket{2} = \ket{0}$.
When $p>14(1-q)/95$ and $q>14(1-p)/95$, the two inequalities (\ref{ex4-2pro-1}) in Appendix III are violated and $\rho$ is genuinely $3$-partite entangled. The results are compared with the ones in \cite{Hong_2021},  { which are the best results so far} in FIG. \ref{fig:compare}.

For the $W$ states mixed with white noise in Example 1 and Example 2, our inequalities perform also better compared with the results in \cite{Hong_2021}.

{\bf{Example 3.}} Consider the $5$-qubit mixed states,
\begin{equation*}
    \rho_{CL_5^s} = (1-t) \frac{\mathbb{I}}{2^5} + t \ket{CL_5^s} \bra{CL_5^s},
\end{equation*}
where $\ket{CL_5^s} = \frac{1}{2} ( \ket{00000} + \ket{01111} + \ket{10011} + \ket{11100})$.

The results are illustrated in Fig. \ref{fig:cl5}, see also the details given in Appendix III. The results in \cite{Zhou} show that $\rho_{CL_5^s}$ is not $5$-partitionable ($4$- or $3$-partitionable) when $t>{7}/{13}$ {while the ones by our criteria are $t>{1}/{9}$ ($t>{5}/{29}$ or $t>{9}/{25}$)}, and
$\rho_{CL_5^s}$ is $2$-partionable or genuine multipartite entangled when $t>{9}/{13}$, {while $t>{13}/{29}$ by our method.}
\begin{figure}
\includegraphics[width=1.0\columnwidth]{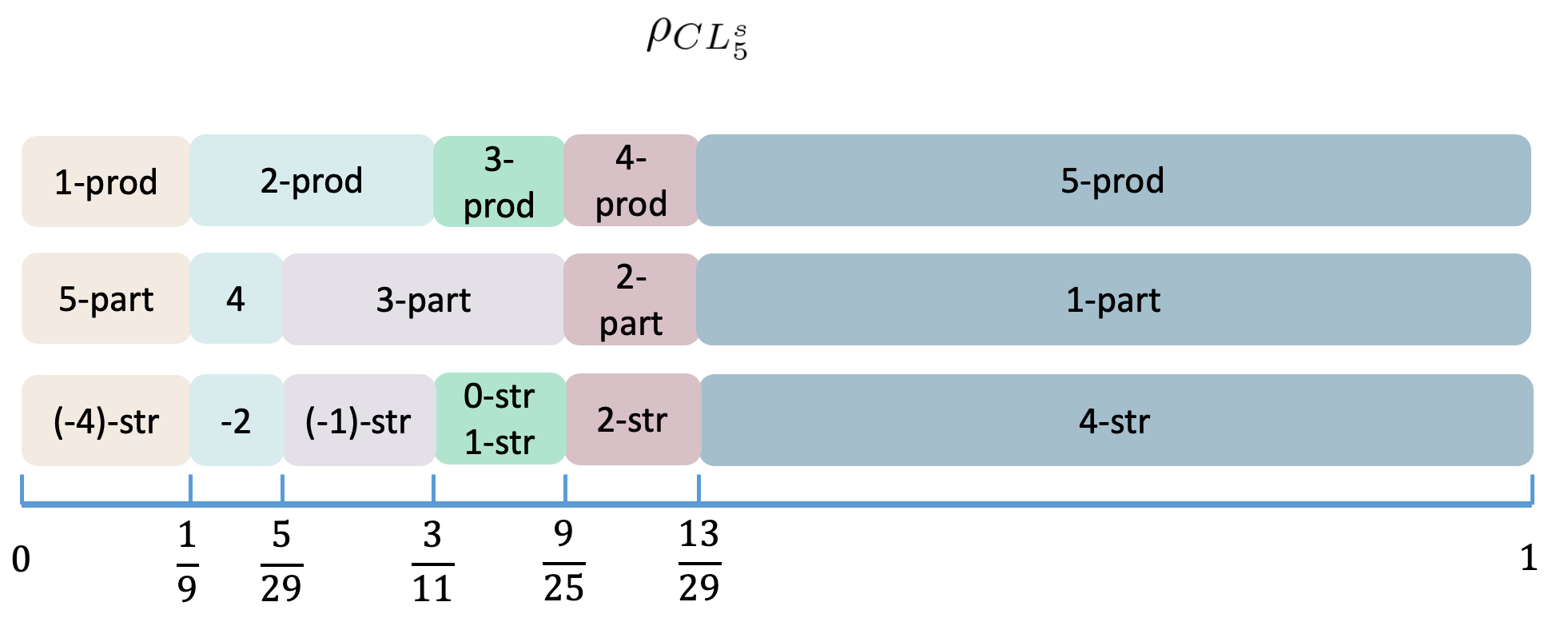}
\caption{\label{fig:cl5} Illustration of the results obtained by using our criteria in detecting the entanglement depth and intactness of $\rho_{CL_5^s}$.}
\end{figure}

{\bf{Example 4.}} Consider the $6$-qubit mixed states,
\begin{equation*}
    \rho_{GHZ_6} = (1-t) \frac{\mathbb{I}}{2^6} + t \ket{GHZ_6} \bra{GHZ_6},
\end{equation*}
where $\ket{GHZ_6} = \frac{1}{\sqrt{2}} ( \ket{000000} + \ket{111111} )$.

In this case, the witness in \cite{Lu} can detect $2$-partitional entanglement only when $t>{32}/{63}$, the genuine entanglment only when $p>{31}/{47}$, $2$ ($3$,$4$,$5$ and $6$)-partite entanglement when $t>0.366$ ($t>0.448$,$p>0.522$,$0.556$ and $p>0.649$).  The witness in \cite{Zhou} can only detect the genuine mutipartite entanglement of $\rho_{GHZ_6}$ when $p>{31}/{47}$. While our criteria can detect the genuine entanglement when $p>{31}/{63}$, and 2(3,4, and 5)-partite entanglement when $p>{1}/{33}$ ($p>{5}/{37}$, $p>{15}/{47}$, and $p>{25}/{57}$). The results are compared with the ones in \cite{Hong_2021}, which are the best known results so far, see Fig. \ref{fig:ghz-6} in page 8 and further details are given in Appendix III. Our algorithm can detect more entangled regions compared with the previous ones.
\begin{figure*}
\includegraphics[width=2.0\columnwidth]{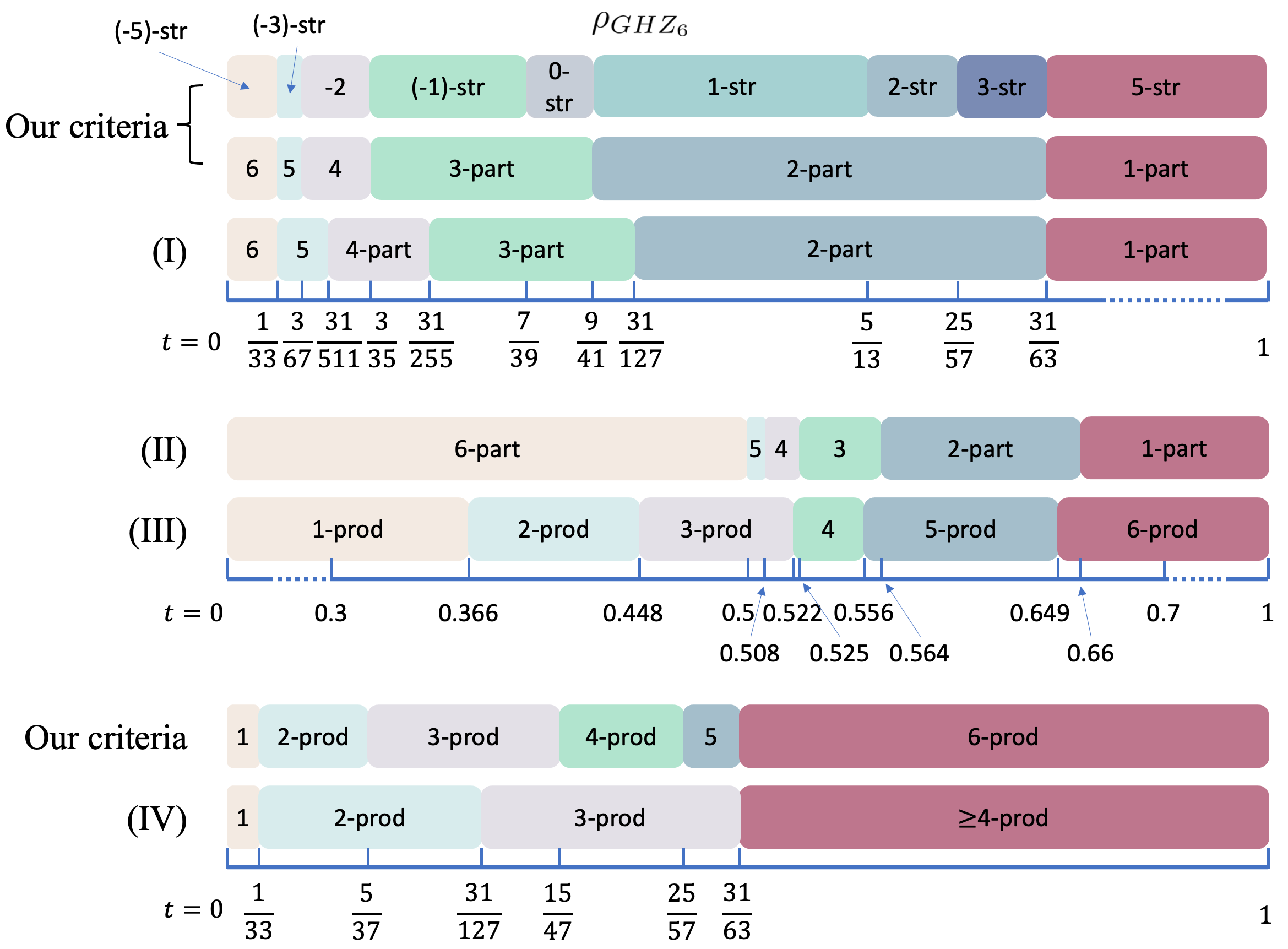}
\caption{\label{fig:ghz-6} Illustration of the results obtained by using our criteria in detecting the  $k$-partitionability and $k$-producibility of $\rho_{GHZ_6}$. In the figure, (I) showcases the results derived from Theorem 3 of Ref. \cite{Hong_2021}, (IV) presents the results from Theorem 1 of Ref. \cite{Hong_2021}, (II) demonstrates the results based on Equation A25 of Ref. \cite{Lu}, and (III) illustrates the results based on Equation A31 of Ref. \cite{Lu}.
}
\end{figure*}

Our criteria based on the hyperplane can detect more $k$-partionable, $k$-producible and $k$-stretchable states compared with the previous analytical results, and can be implemented experimentally by measuring local observables. Take $6$-qubit GHZ state mixed with white noise, $W^P$
can be realized by using $32$ local observables.
To detect the $k$-nonseperable states, the observable $W_{k-\text{part}}$ can be realized using  $4$ $(34,14,14,4)$ local observables for $k=2$ $(3,4,5,6)$, respectively. To detect the $k$-partite states, the observable $W_{k-\text{prod}}$ can be realized by using  $4 (12,32,16,4)$ local observables for $k=2$ $(3,4,5,6)$, respectively.

\section{Conclusion}
Based on the property that all states with the same entanglement structre constitute a convex set, we have designed an algorithm by using SDP and the gradient descent method to appromixmate the convex set from the inside of it. {{The gradient descent method optimizes the initial points for the SDP, thus the probability to obtain the global optimum is increased.}}
And we have also constructed the criteria from the outer polytopes of the convex set to characterize the multipartite entanglement structure. These criteria can be implemented through local observables and the number of the ones grows linearly with the system size. {{The criteria can be optimized further by optimizing the selection of the off-diagonal elements and diagnoal elements of the density matrices.}} The algorithms and the criteria can characterize the multipartite entanglement structure better compared with existing criteria.

\medskip
\noindent{\bf ACKNOWLEDGEMENTS}\, \, This work is supported by the Fundamental Research Funds for the Central Universities; the National Natural Science Foundation of China (NSFC) under Grants  12071179, 12305007, 12371132, 12075159 and 12171044; Anhui Provincial Natural Science Foundation (Grant No.
2308085QA29); the Alexander von Humboldt Foundation;  Natural Science Foundation of Shanghai (Grant No. 20ZR1426400); Shenzhen Institute for Quantum Science and Engineering, Southern University of Science and Technology (Grant Nos. SIQSE202005); the Academician Innovation Platform of Hainan Province.

\clearpage

\section{Appendix I}

The proof of Proposition 1 is given as follows.
\begin{widetext}
{\bf{Proof of Proposition \ref{prop:1}}}
\begin{align*}
\begin{aligned}
&|\bra{m}\psi\rangle\langle\psi\ket{n}| = \sqrt{\bra{m}\psi\rangle\langle\psi\ket{n} \bra{n}\psi\rangle\langle\psi\ket{m}} \\
=&\sqrt{(\bra{m_{X_i}} \bra{m_{\bar{X}_i}}) \ket{\psi_{X_i}}  \ket{\psi_{\bar{X}_i}}} \times \sqrt{ \bra{\psi_{X_i}}  \bra{\psi_{\bar{X}_i}} (\ket{n_{X_i}} \ket{n_{\bar{X}_i}}) } \times \sqrt{(\bra{n_{X_i}} \bra{n_{\bar{X}_i}}) \ket{\psi_{X_i}}  \ket{\psi_{\bar{X}_i}} } \times \sqrt{ \bra{\psi_{X_i}}  \bra{\psi_{\bar{X}_i}} ( \ket{m_{X_i}} \ket{m_{\bar{X}_i}}) }\\
=&\sqrt{ \bra{m_{X_i}} \psi_{X_i} \rangle \bra{m_{\bar{X}_i}} \psi_{\bar{X}_i} \rangle } \times \sqrt{\bra{\psi_{X_i}}  n_{X_i} \rangle \bra{\psi_{\bar{X}_i}} n_{\bar{X}_i} \rangle } \times \sqrt{\bra{n_{X_i}} \psi_{X_i} \rangle \bra{n_{\bar{X}_i}} \psi_{\bar{X}_i} \rangle } \times \sqrt{ \bra{\psi_{X_i}} m_{X_i} \rangle \bra{\psi_{\bar{X}_i}} m_{\bar{X}_i} \rangle } \\
=&\sqrt{ \bra{m_{X_i}} \psi_{X_i} \rangle \bra{n_{\bar{X}_i}} \psi_{\bar{X}_i} \rangle } \times \sqrt{ \bra{\psi_{X_i}}   m_{X_i} \rangle \bra{\psi_{\bar{X}_i}} n_{\bar{X}_i} \rangle  } \times \sqrt{ \bra{n_{X_i}} \psi_{X_i} \rangle \bra{m_{\bar{X}_i}} \psi_{\bar{X}_i} \rangle } \times \sqrt{\bra{\psi_{X_i}} n_{X_i} \rangle \bra{\psi_{\bar{X}_i}} m_{\bar{X}_i} \rangle } \\
=&\sqrt{(\bra{m_{\bar{X}_i}} \bra{n_{X_i}}) \ket{\psi} \bra{\psi} (\ket{m_{\bar{X}_i}} \ket{n_{X_i}})} \times \sqrt{ (\bra{m_{X_i}} \bra{n_{\bar{X}_i}}) \ket{\psi} \bra{\psi} (\ket{m_{X_i}} \ket{n_{\bar{X}_i}})} \\
\leq&\frac{1}{2} [ (\bra{m_{\bar{X}_i}} \bra{n_{X_i}}) \ket{\psi} \bra{\psi} (\ket{m_{\bar{X}_i}} \ket{n_{X_i}})  + (\bra{m_{X_i}} \bra{n_{\bar{X}_i}}) \ket{\psi} \bra{\psi} (\ket{m_{X_i}} \ket{n_{\bar{X}_i}}) ].
\end{aligned}
\end{align*}
\end{widetext}

\section{Appendix II}

{ In the following part, we illustrate the optimization process for the inequality based on outer polytopes. The left-hand side of the inequality represents the sum of the absolute values of the non-diagonal elements in $P$, while the right-hand side is the inner product of the vectors $\vec{\alpha}$ and $\vec{\beta}$. $\vec{\alpha}$ is derived from the coefficient matrix $M$ by taking the maximum value of per column. The entire optimization process can be viewed as the optimization of a combination of inequalities.

The  optimization strategy adheres to two fundamental principles: (1)  Maximize the number of all zero columns within matrix $ M $. (2) Minimize the  sum of the elements comprising the vector $ \vec{\alpha} $.}

{ For each element $\rho_{m,n}$ within the set $P$, at least one corresponding inequality can be formulated for every partition $\gamma_i$. We denote the set of the inequalities with respect to $\rho_{m,n}$ and $\gamma_i$ as $S_{m,n}^{\gamma_i}.$
%Given our preference for inequalities with a minimal number of diagonal elements on their right-hand side,
We construct a candidate set from the inequalities in $S_{m,n}^{\gamma_i}$ for all $\gamma_i\in \Gamma^{max}$ and given $\rho_{mn}$, which satisfies two conditions: (1) containing  one inequality for every partition, (2) containing as few inequalities as possible. It is acknowledged that there are multiple such sets  that could fulfill these conditions.  Based on the principle of minimizing the number of diagonal elements on the right-hand side of the inequalities, an optimal candidate set $ S_{m,n}$ is selected for each element $\rho_{m,n}$ from the set $ P $.

Then the given partition $\gamma_i$ and  each element $\rho_{m,n}$ and  we select  one inequality  from the set $S_{m,n}^{\gamma_i}\cap S_{m,n}$  and then we add the left and the righ hand sides of all the inequalities respectively  for $m$ and $n,$ the inequality correspoing to the $i$-th row of $M$ is attained.
The selection princile is minimizing the  summuation of the elements that comprise the vector $\vec{\alpha}$, The detais of the selecting are as follows.

There may exist a specific partition \( \gamma \) such that for each part \( X_i \) within \( \gamma \), it holds that \( \ket{m_{X_i}} = \ket{n_{X_i}} \). This implies that for \( \rho_{m,n} \), the corresponding inequality associated with partition \( \gamma \) is \( |\rho_{m,n}| \leq (\rho_{m,m} + \rho_{n,n})/2 \), which is  trivial from the positive definiteness of the density matrices.
For each partition $\gamma_i$ in $\Gamma^{\max}$, all trivial inequalities associated with $\gamma_i$ can be determined. By summing the left and the right hand sides of these trivial inequalities respectively, we can obtain the corresponding trivial inequality for $\gamma_i$.

The right-hand side of the trivial inequalities for all partitions in $\Gamma^{max}$ can also be expressed as the inner product of a matrix $ M' $ with the vector $ \vec{\beta} $. Subsequently, by taking the maximum value across the columns of $ M' $, we obtain the vector $\vec{\eta}$. The maximum values in all columns of matrix $M$, which correspond to all trivial inequalities, are determined by $\vec{\eta}$. The vector $ \vec{\eta} $ can be considered as the zeroth row of the matrix $ M $. Then, when constructing a certain row of the matrix $ M $, if the value in the zeroth row of a certain column is not equal to zero, and the value in that row and column of $M$ is smaller than the value in the zeroth row, it may be beneficial to adjust the selection of inequalities such that the value in that row and column approaches the value in the zeroth row. This approach aligns well with our guiding principle for selection. This process involves exact matching, for which we have utilized the Dancing Links algorithm \cite{knuth2000dancing}. An illustration of this approach will be provided in the forthcoming details of Example 1. All of the aforementioned constitutes our comprehensive optimization strategies and techniques.}

\section{Appendix III}
\begin{figure}
\includegraphics[width=1.0\columnwidth]{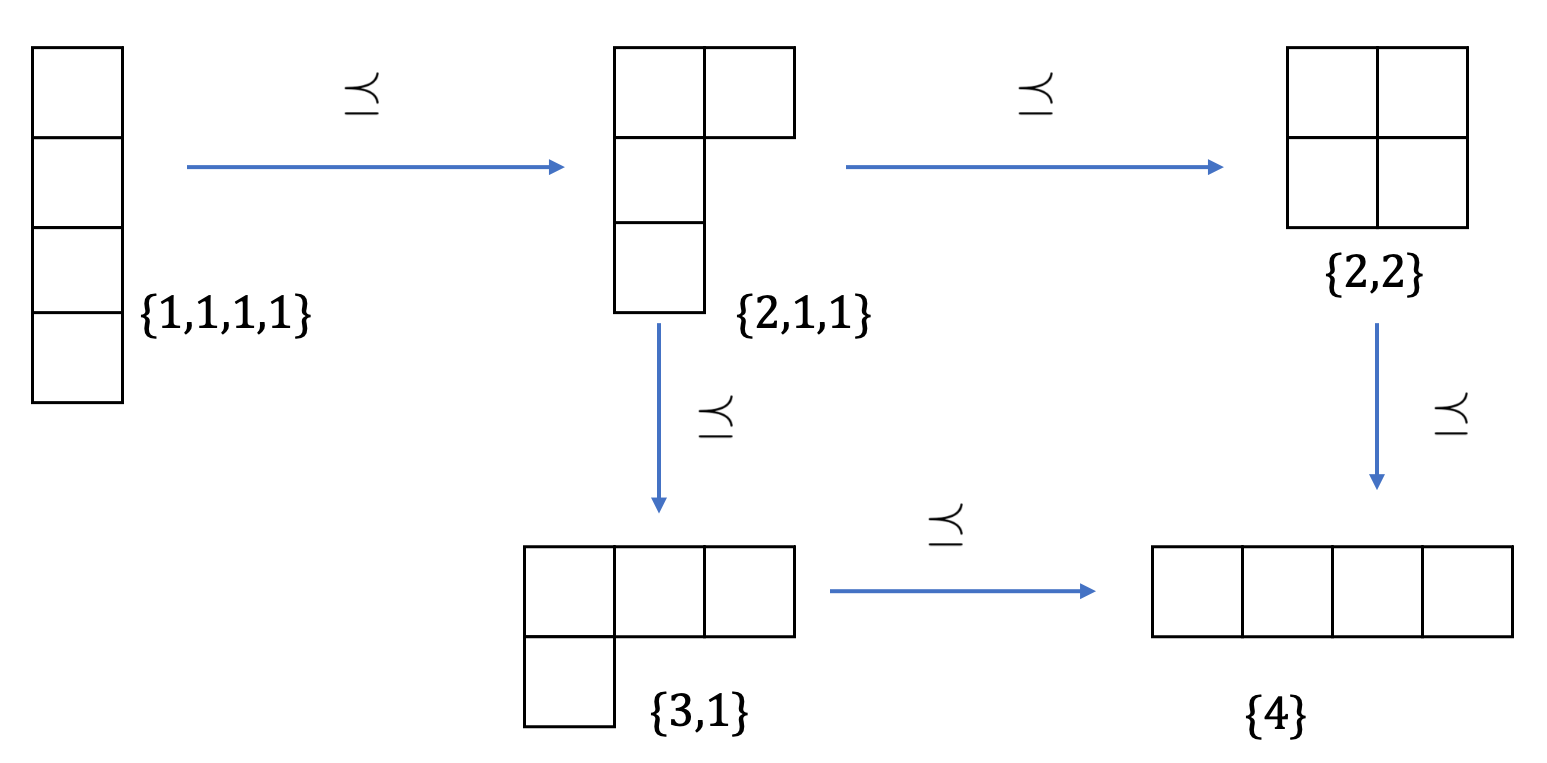}
\caption{\label{fig:partition-type} The types of all partitions of the set $ \{1, 2, 3, 4\} $ are represented by Young diagrams, along with their poset relationships.}
\end{figure}

{{In the following section, we give the details to detect the entanglement structure of the examples using the criteria based on the outerplane.
To represent the partitions for short in the examples, we give the definitions of the type of the partitions firstly. The type of the partition $\gamma$ (the integer partition of $N$) is denoted as $\hat{\gamma}=\{|X_1|,\cdots, |X_{|\gamma|}| \}$, where $X_i$ is the part of $\gamma$. Disregarding the sequence of elements, the type of the partition $ \gamma $ is uniquely identified. $\hat{\gamma}$ is the refinement of $ \hat{\xi}$ if $\gamma$ is the refinement of $\xi$. as depicted in Fig. \ref{fig:partition-type}, Young diagrams provide a graphical representation for the types of partitions. If $\hat{\gamma}$ is the refinement of $ \hat{\xi}$, then for each partition $\gamma_1$ in $\hat{\gamma}$, there exists a partition $\xi_1$ in $\hat{\xi}$ such that $\gamma_1 \preceq \xi_1$.}} 

{{For example, as depicted in Fig. \ref{fig:partition-type}, when we consider the set of partition types for $ \Gamma_{2-\text{part}} $ of $ \{1, 2, 3, 4\} $, we find that it consists of $ \{\{1, 1, 1, 1\}, \{2, 1, 1\}, \{2, 2\}, \{3, 1\}\} $. Among these, the maximal elements within the set of partition types are $ \{\{2, 2\}, \{3, 1\}\} $. Consequently, $ \Gamma_{2-\text{part}}^{\max} $ encompasses all partitions whose types are either $ \{2, 2\} $ or $ \{3, 1\} $, reflecting the partitions that achieve the greatest refinement within the poset defined by $ \Gamma_{2-\text{part}} $ .}} 

Then we illustrated the details of the Examples in E of Section III.

\noindent{\bf The details of Example 1:}

{We select the set of non-diagonal elements $P = \{ \rho_{2,3}, \rho_{2,5}, \rho_{2,9}, \rho_{3,5}, \rho_{3,9}, \rho_{5,9} \}$.}

The sets of maximal elements of all $k$-stretchable partitions of $\{1,2,3,4\}$ are
{\begin{eqnarray*}
    \Gamma_{(-1)-\text{str}}^{\max}&=&\{12|3|4, 13|2|4, 14|2|3, 23|1|4, 24|1|3, 34|1|2\},\\
    \Gamma_{0-\text{str}}^{\max}&=&\{12|34, 13|24, 14|23\}, \\
    \Gamma_{1-\text{str}}^{\max}&=&\Gamma_{0-\text{str}}^{\max} \bigcup \{123|4, 124|3, 134|2, 234|1\}.
\end{eqnarray*}}

{If $\rho$ is $3$-partitionable or $(-1)$-stretchable, the trivial inequality corresponding to $12|3|4$ is $|\rho_{5,9}| \le (\rho_{5,5} + \rho_{9,9}) / 2$. Additionally, the trivial inequality corresponding to $12|3|4$ is $|\rho_{2,3}| \le (\rho_{2,2}+\rho_{3,3}) / 2$. As a result, the vector $\vec{\eta} = [ \eta_1, \eta_2, ..., \eta_{16} ]$ can be defined with $\eta_2 = \eta_3 = \eta_5 = \eta_9 = 1/2$. When constructing the inequality corresponding to $1|2|34$, if we select $|\rho_{5,9}| \le (\rho_{1,1} + \rho_{13,13}) / 2$, the coefficient of $\rho_{1,1}$ in the final inequality becomes $2.5$. However, if we select $|\rho_{5,9}| \le (\rho_{5,5} + \rho_{9,9}) / 2$, it appears to add a trivial inequality, but the coefficient of $\rho_{1,1}$ in the final inequality becomes $2$, with no change in the coefficients of other terms. This is the optimization method for adjusting inequality selection introduced in Appendix II.

Therefore, if $\rho$ is $3$-partitionable or $(-1)$-stretchable, the partition $12|3|4$ and $34|1|2$ correspond to the following inequality:
\begin{align}\label{w-4-inq-base-1}
\begin{aligned}
\sum_{\rho_{m,n} \in P} |\rho_{m,n}| \leq& 2\rho_{1,1} + \frac{1}{2}(\rho_{6,6}+\rho_{7,7}+\rho_{10,10}+\rho_{11,11}\\
&+\rho_{2,2}+\rho_{3,3}+\rho_{5,5}+\rho_{9,9}),
\end{aligned}
\end{align}
and the partition $13|2|4$ and $24|1|3$ correspond to the following inequality:
\begin{align}\label{w-4-inq-base-2}
\begin{aligned}
\sum_{\rho_{m,n} \in P} |\rho_{m,n}| \leq& 2\rho_{1,1} + \frac{1}{2}(\rho_{4,4}+\rho_{7,7}+\rho_{10,10}+\rho_{13,13}\\
&+\rho_{2,2}+\rho_{3,3}+\rho_{5,5}+\rho_{9,9}),
\end{aligned}
\end{align}
and the partition $14|2|3$ and $23|1|4$ correspond to the following inequality:
\begin{align}\label{w-4-inq-base-3}
\begin{aligned}
\sum_{\rho_{m,n} \in P} |\rho_{m,n}| \leq& 2\rho_{1,1} + \frac{1}{2}(\rho_{4,4}+\rho_{6,6}+\rho_{11,11}+\rho_{13,13}\\
&+\rho_{2,2}+\rho_{3,3}+\rho_{5,5}+\rho_{9,9}).
\end{aligned}
\end{align}
Thus one of the criteria is
\begin{align}\label{w-4-inq-base}
\begin{aligned}
\sum_{\rho_{m,n} \in P} |\rho_{m,n}| \leq& 2\rho_{1,1} + \frac{1}{2}(\rho_{4,4}+\rho_{6,6}+\rho_{7,7}+\rho_{10,10}+\rho_{11,11}\\
&+\rho_{13,13}+\rho_{2,2}+\rho_{3,3}+\rho_{5,9}+\rho_{9,9}).
\end{aligned}
\end{align}
}

{If $\rho$ is $2$-producible or $0$-stretchable, the partition $12|34$ corresponds to the inequality (\ref{w-4-inq-base-1}), and the partition $13|24$ corresponds to the inequality (\ref{w-4-inq-base-2}), and the partition $14|23$ corresponds to the inequality (\ref{w-4-inq-base-3}). Thus one of the criteria is the inequality (\ref{w-4-inq-base}).

If $\rho$ is $3$-producible, $2$-partitionable, or $1$-stretchable, one of the criteria is
\begin{align}
\begin{aligned}
\sum_{\rho_{m,n} \in P} |\rho_{m,n}| \leq& 2\rho_{1,1} + \frac{1}{2}(\rho_{4,4}+\rho_{6,6}+\rho_{7,7}+\rho_{10,10}+\rho_{11,11}\\
&+\rho_{13,13})+(\rho_{2,2}+\rho_{3,3}+\rho_{5,9}+\rho_{9,9}).
\end{aligned}
\end{align}}

The observable $W_P$ and $W_{k-\text{prod}}$ in the left and right hand sides of the inequalies in this example can be expressed as
\begin{align}
\begin{aligned}
&W_P = \sum_{1 \le i < j \le n}[\bigotimes_{k\neq i,j} Z^{+}_{k}]\otimes H_{i,j},\\
&W_{2-\text{prod}}=2\Delta_1+\frac{1}{2}(\Delta_2+\Delta_3)\\
&W_{3-\text{prod}}=2\Delta_1+\Delta_2+\frac{1}{2}\Delta_3
\end{aligned}
\end{align}
with
\begin{align*}
\begin{aligned}
&Z_i^{+} = \frac{(\mathbb{I}_2 + \sigma^{(i)}_z)}{2}, Z_k^{-} = \frac{(\mathbb{I}_2 - \sigma^{(i)}_z)}{2},\\
&H_i^{+} = \ket{0}_i\bra{1}, H_i^{-} = \ket{1}_i\bra{0}\\
&H_{i,j} = H_i^{+} \otimes H_j^{-} + H_i^{-} \otimes H_j^{+}\\
&\Delta_1=\bigotimes\limits_{i=1}^4 Z_k^{+}\\
&\Delta_2=\sum\limits_{i=1}^4 Z_i^{-}\otimes [\bigotimes\limits_{j=1,j\neq i}^4 Z_j^{+}]\\
&\Delta_3=\sum\limits_{1\le i_1< i_2\le 4} Z_{i_1}^+\otimes
Z_{i_2}^+\otimes [\bigotimes\limits_{j \neq i_1,i_2} Z_j^-]
\end{aligned}
\end{align*}
and $\sigma_z^{(i)}$ being the pauli matrice of the $i$-th system.

\medskip
\noindent{\bf The details of Example 2.}

{We select $P_1=\{$ $\rho_{2,3}$, $\rho_{2,4}$, $\rho_{2,7}$, $\rho_{2,10}$, $\rho_{2,19}$, $\rho_{2,28}$, $\rho_{2,55}$, $\rho_{3,4}$, $\rho_{3,7}$, $\rho_{3,10}$, $\rho_{3,19}$, $\rho_{3,28}$, $\rho_{3,55}$, $\rho_{4,7}$, $\rho_{4,10}$, $\rho_{4,19}$, $\rho_{4,28}$, $\rho_{4,55}$, $\rho_{7,10}$, $\rho_{7,19}$, $\rho_{7,28}$, $\rho_{7,55}$, $\rho_{10,19}$, $\rho_{10,28}$, $\rho_{10,55}$, $\rho_{19,28}$, $\rho_{19,55}$, $\rho_{28,55}$ $\}$, and $P_2=\{$ $\rho_{14,32}$, $\rho_{14,38}$, $\rho_{14,40}$, $\rho_{14,42}$, $\rho_{14,44}$, $\rho_{14,50}$, $\rho_{14,68}$, $\rho_{32,38}$, $\rho_{32,40}$, $\rho_{32,42}$, $\rho_{32,44}$, $\rho_{32,50}$, $\rho_{32,68}$, $\rho_{38,40}$, $\rho_{38,42}$, $\rho_{38,44}$, $\rho_{38,50}$, $\rho_{38,68}$, $\rho_{40,42}$, $\rho_{40,44}$, $\rho_{40,50}$, $\rho_{40,68}$, $\rho_{42,44}$, $\rho_{42,50}$, $\rho_{42,68}$, $\rho_{44,50}$, $\rho_{44,68}$, $\rho_{50,68}$ $\}$.}

If $\rho$ is $2$-producible, then
\begin{eqnarray}
\label{ex4-2pro-1}
&&{\sum\limits_{\rho_{m,n} \in P_i}|\rho_{m,n}|} \le \textrm{Tr}[W_{2-\text{prod}}^{P_i} \cdot \rho],
\end{eqnarray}

where $W_{2-\text{prod}}^{P_i}=\textrm{diag}\{a^i_{1},\cdots\}$ $(i=1,2)$ is the diagonal matrix with $a^1_1=8,$
$a^1_2=a^1_3=a^1_4=a^1_7=a^1_{10}=a^1_{19}=a^1_{28}=a^1_{55}=1$
and $a^1_5=a^1_6=a^1_8=a^1_9=a^1_{11}=a^1_{12}=a^1_{13}=a^1_{16}=a^1_{20}=a^1_{21}=a^1_{22}=a^1_{25}=a^1_{29}=a^1_{30}=a^1_{31}=a^1_{34}=a^1_{37}=a^1_{46}=a^1_{56}=a^1_{57}=a^1_{58}=a^1_{61}=a^1_{64}=a^1_{73}=\frac{1}{2}.$
and
$a^2_{41}=8,$
$a^2_{14}=a^2_{32}=a^2_{38}=a^2_{40}=a^2_{42}=a^2_{44}=a^2_{50}=a^2_{68}=1$
and $a^2_5=a^2_{11}=a^2_{13}=a^2_{15}=a^2_{17}=a^2_{23}=a^2_{28}=a^2_{31}=a^2_{33}=a^2_{35}=a^2_{37}=a^2_{39}=a^2_{43}=a^2_{45}=a^2_{47}=a^2_{49}=a^2_{51}=a^2_{53}=a^2_{59}=a^2_{65}=a^2_{67}=a^2_{69}=a^2_{71}=a^2_{77}=\frac{1}{2}.$

The observable $W_{P_1}$, $W_{P_2}$, $W_{2-\text{prod}}^{P_1}$ and $W_{2-\text{prod}}^{P_2}$ in the left and right hand sides of the inequalies in this example can be expressed as

\begin{align}
\begin{aligned}
&W_{P_1} = \sum_{1 \le i < j \le 4}[(\bigotimes_{k\neq i,j} Z^{0}_{k})\otimes (H_{i,j} + J_{i,j} + \Delta_{i,j}^1 + \Delta_{j,i}^1 )] \\
&W_{P_2} = \sum_{1 \le i < j \le 4}[(\bigotimes_{k\neq i,j} Z^{1}_{k})\otimes ( H_{i,j} +  K_{i,j} + \Delta_{i,j}^2 + \Delta_{j,i}^2 )] \\
&W_{2-\text{prod}}^{P_1} = 8 \cdot(\bigotimes\limits_{i=1}^4 Z_i^{0}) + \frac{1}{2} \sum_{1 \le i < j \le 4} [ ( \sum_{a,b \in \{ 1, 2 \}} Z_i^{a} \otimes Z_j^{b}) \\
&\otimes (\bigotimes_{k \neq i, j} Z_k^{0}) ] + \sum_{1 \le i \le 4}[(\bigotimes_{k\neq i} Z^{0}_{k}) \otimes ( Z^{1}_{i} + Z^{2}_{i} )] \\
&W_{2-\text{prod}}^{P_2} = 8 \cdot(\bigotimes\limits_{i=1}^4 Z_i^{1}) + \frac{1}{2} \sum_{1 \le i < j \le 4} [ (\sum_{a,b \in \{ 0, 2 \}} Z_i^{a} \otimes Z_j^{b}) \\
&\otimes (\bigotimes_{k \neq i, j} Z_k^{1}) ] + \sum_{1 \le i \le 4}[(\bigotimes_{k\neq i} Z^{1}_{k}) \otimes ( Z^{0}_{i} + Z^{2}_{i} )]
\end{aligned}
\end{align}
with
\begin{align}
\begin{aligned}
&Z_i^{0} = \ket{0}_i\bra{0}, Z_i^{1} = \ket{1}_i\bra{1}, Z_i^{2} = \ket{2}_i\bra{2}\\
&H_i^{+} = \ket{0}_i\bra{1}, H_i^{-} = \ket{1}_i\bra{0}\\
&J_i^{+} = \ket{0}_i\bra{2}, J_i^{-} = \ket{2}_i\bra{0}\\
&K_i^{+} = \ket{1}_i\bra{2}, K_i^{-} = \ket{2}_i\bra{1}\\
&H_{i,j} = H_i^{+} \otimes H_j^{-} + H_i^{-} \otimes H_j^{+}\\
&J_{i,j} = J_i^{+} \otimes J_j^{-} + J_i^{-} \otimes J_j^{+}\\
&K_{i,j} = K_i^{+} \otimes K_j^{-} + K_i^{-} \otimes K_j^{+}\\
&\Delta_{i,j}^1 = H_i^{+} \otimes J_j^{-} + H_i^{-} \otimes J_j^{+}\\
&\Delta_{i,j}^2 = H_i^{+} \otimes K_j^{-} + H_i^{-} \otimes K_j^{+}\\
\end{aligned}
\end{align}

\medskip
\noindent{\bf The details of Example 3.}

Unless otherwise specified, we select {$P=\{ \rho_{1,16}, \rho_{1,20}, \rho_{1,29}, \rho_{16,20}, \rho_{16,29}, \rho_{20,29} \}$.}

If $\rho$ is $(-2)$-stretchable or $4$-partitionable, $\Gamma_{4-\text{part}}^{\max}$ is a set of partitions having the type $\{ 2, 1, 1, 1 \}$. By the greedy algorithm, one of the criteria is
\begin{eqnarray}
\label{ex4-4par}
{\sum_{\rho_{m,n} \in P}|\rho_{m,n}|} \le \textrm{Tr}[W_{4-\text{part}} \cdot \rho],
\end{eqnarray}
where $W_{4-\text{part}}=\textrm{diag}\{a_1,\cdots\}$ is the diagonal matrix with $a_{2} = a_{3} = a_{4} = a_{5} = a_{8} = a_{9} = a_{12} = a_{13} = a_{14} = a_{15} = a_{17} = a_{18} = a_{19} = a_{21} = a_{24} = a_{25} = a_{28} = a_{30} = a_{31} = a_{32} = \frac{1}{2}$.

If $\rho$ is $(-1)$-stretchable or $2$-producible, $\Gamma_{2-\text{prod}}^{\max}$ is a set of partitions having the type $\{ 2, 2, 1 \}$. We select {$P_2=\{ \rho_{1,16} \}$.  By the codes in \cite{Wu},} one of the criteria is \begin{equation}
\label{ex4-2pro}
{|\rho_{1,16}|} \le \textrm{Tr}[W_{2-\text{prod}} \cdot \rho],
\end{equation}
where $W_{2-\text{prod}}=\textrm{diag}\{a_1,\cdots\}$ is the diagonal matrix with $a_{4} = a_{6} = a_{7} = a_{10} = a_{11} = a_{13} = \frac{1}{2}$.

If $\rho$ is $0$-stretchable or $1$-stretchable or $3$-producible or $3$-partitionable, one of the criteria is
\begin{equation}
\label{ex4-3pro}
{\sum_{\rho_{m,n} \in P}|\rho_{m,n}|} \le \textrm{Tr}[W_{3-\text{part}} \cdot \rho],
\end{equation}
where $W_{3-\text{part}}=\textrm{diag}\{a_1,\cdots\}$ is the diagonal matrix with $a_{4} = a_{13} = a_{17} = a_{32} = 1$, and $a_{1} = a_{2} = a_{3} = a_{5} = a_{6} = a_{7} = a_{8} = a_{9} = a_{10} = a_{11} = a_{12} = a_{14} = a_{15} = a_{16} = a_{18} = a_{19} = a_{20} = a_{21} = a_{22} = a_{23} = a_{24} = a_{25} = a_{26} = a_{27} = a_{28} = a_{29} = a_{30} = a_{31} = \frac{1}{2}$.

If $\rho$ is $2$-stretchable or $2$-partitionable or $4$-producible, one of the criteria is
\begin{eqnarray}
\label{ex4-4pro}
{\sum_{\rho_{m,n} \in P}|\rho_{m,n}|} \le \textrm{Tr}[W_{4-\text{prod}} \cdot \rho],
\end{eqnarray}
where $W_{4-\text{prod}}=\textrm{diag}\{a_1,\cdots\}$ is the diagonal matrix with $a_{2} = a_{3} = a_{4} = a_{5} = a_{8} = a_{9} = a_{12} = a_{13} = a_{14} = a_{15} = a_{17} = a_{18} = a_{19} = a_{21} = a_{24} = a_{25} = a_{28} = a_{30} = a_{31} = a_{32} = 1$, and $a_{1} = a_{6} = a_{7} = a_{10} = a_{11} = a_{16} = a_{20} = a_{22} = a_{23} = a_{26} = a_{27} = a_{29} = \frac{1}{2}$.

The observables $W_P$ and $W_{k-\text{prod}},$ $W_{k-\text{part}}$ and $W_{k-\text{str}}$ are expressed as follows
\begin{align}
\begin{aligned}
&W_P = Z_1^{+} \otimes (\bigotimes\limits_{i=2}^5 H_i^{+}) + Z_2^{+} \otimes Z_3^{+} \otimes H_1^{+} \otimes H_4^{+} \otimes H_5^{+} \\
&+ Z_4^{+} \otimes Z_5^{+} \otimes H_1^{+} \otimes H_2^{+} \otimes H_3^{+} + Z_2^{-} \otimes Z_3^{-} \otimes H_1^{+} \\
&\otimes H_4^{-} \otimes H_5^{-} + Z_4^{-} \otimes Z_5^{-} \otimes H_1^{+} \otimes H_2^{-} \otimes H_3^{-} \\
&+ Z_1^{-} \otimes H_2^{+} \otimes H_3^{+} \otimes H_4^{-} \otimes H_5^{-} \\
&W_{P_2} = Z_1^{+} \otimes (\bigotimes\limits_{i=2}^5 H_i^{+}) \\
&W_{4-\text{part}} = \frac{\mathbb{I}_{32} - \Delta_1}{2} \\
&W_{2-\text{prod}} = \frac{1}{2} Z_1^{+} \otimes \sum_{2 \le i < j \le 5} [Z_i^{+} \otimes Z_j^{+} \otimes (\bigotimes_{\substack{2 \le k \le 5,\\ k \neq i,j}} Z_k^{-})] \\
&W_{3-\text{part}} = \frac{\mathbb{I}}{2} + \frac{\Delta_2}{2} \\
&W_{4-\text{prod}} = \mathbb{I} - \frac{\Delta_1}{2}
\end{aligned}
\end{align}
with
\begin{align}
\begin{aligned}
&\Delta_1 = (Z_1^{-} + Z_1^{+}) \otimes [\sum_{\substack{2 \le i \le 3,\\ 4 \le j \le 5}} Z_i^{+} \otimes Z_j^{+} \otimes (\bigotimes_{\substack{2 \le k \le 5,\\ k \neq i,j}} Z_k^{-} )] \\
&+ \bigotimes_{i=1}^5 Z_i^{-} + Z_1^{-} \otimes (\bigotimes_{i=2}^5 Z_i^{+}) + Z_1^{+} \otimes ( Z_2^{-} \otimes Z_3^{-} \otimes Z_4^{+} \\
&\otimes Z_5^{+} + Z_2^{+} \otimes Z_3^{+} \otimes Z_4^{-} \otimes Z_5^{-}) \\
&\Delta_2 = Z_1^{-} \otimes ( Z_2^{-} \otimes Z_3^{-} \otimes Z_4^{+} \otimes Z_5^{+} + Z_2^{+} \otimes Z_3^{+} \otimes Z_4^{-}\\
&\otimes Z_5^{-} + Z_1^{+} \otimes (\bigotimes_{i=2}^5 Z_i^{-} + \bigotimes_{i=2}^5 Z_i^{+} )
\end{aligned}
\end{align}

\medskip
\noindent{\bf The details of Example 4.}

Unless otherwise specified, we select {$P=\{ \rho_{1, 64} \}$}.

Different from other examples, we fix the right-hand side of the inequality and seek a larger left-hand side, the details are listed as follows.

If $\rho$ is $1$-producible, one of the criteria is
\begin{eqnarray}
\label{ex3-6part}
    {|\rho_{1, 64}| }\le \textrm{Tr}[W_{1-\text{prod}} \cdot \rho],
\end{eqnarray}
where $W_{1-\text{prod}}=\textrm{diag}\{a_1,\cdots, a_{64}\}$ is the diagonal matrix with $a_{2} = a_{63} = \frac{1}{2}$.

If $\rho$ is $5$-partitionable or $(-3)$-stretchable, $\Gamma_{5-\text{part}}^{\max}$ is a set of partitions of type $\{ 2, 1, 1, 1, 1 \}$. Consider six inequalities derived from Proposition \ref{prop:1}, corresponding to the parts $\{1\}$, $\{2\}$, $\{3\}$, $\{4\}$, $\{5\}$, and $\{6\}$, respectively. {Given each partition in the current set includes four parts, summing the corresponding four inequalities yields the inequality for that partition}. Thus the criterion is
\begin{eqnarray}
\label{ex3-5part}
{4 |\rho_{1, 64}| }\le \textrm{Tr}[W_{5-\text{part}} \cdot \rho],
\end{eqnarray}
where $W_{5-\text{part}}=\textrm{diag}\{a_1,\cdots\}$ is the diagonal matrix with $a_{2} = a_{3} = a_{5} = a_{9} = a_{17} = a_{32} = a_{33} = a_{48} = a_{56} = a_{60} = a_{62} = a_{63} = \frac{1}{2}$.

If $\rho$ is $4$-partitionable, $\Gamma_{4-\text{part}}^{\max}$ is a set of partitions of type $\{ 2, 2, 1, 1 \}$ and $\{ 3, 1, 1, 1 \}$. {Given each partition in the current set includes two parts, with each part containing one subsystem}, the criterion is
\begin{eqnarray}
\label{ex3-4part}
{2 |\rho_{1, 64}|} \le \textrm{Tr}[W_{4-\text{part}} \cdot \rho],
\end{eqnarray}
where $W_{4-\text{part}}=W_{5-\text{part}}$.

If $\rho$ is $(-2)$-stretchable, $\Gamma_{(-2)-\text{str}}^{\max}$ is a set of partitions of type $\{ 2, 2, 1, 1 \}$. Therefore, the criterion is given by the inequality (\ref{ex3-4part}).

If $\rho$ is $(-1)$-stretchable, $\Gamma_{(-1)-\text{str}}^{\max}$ is a set of partitions of types $\{ 2, 2, 2 \}$ and $\{ 3, 1, 1, 1 \}$. {We consider 15 inequalities derived from Proposition \ref{prop:1}, each corresponding to a part that includes two subsystems. For each partition in the set of partitions of type $\{ 2, 2, 2 \}$, which includes three parts each containing two subsystems, the inequality for that partition is obtained by summing the corresponding three inequalities.
Similarly, we consider six inequalities, each corresponding to a part that contains one subsystem. For each partition in the set of partitions of type $\{ 3, 1, 1, 1 \}$, which includes three parts each containing one subsystem, the inequality for that partition is obtained by summing the corresponding three inequalities.}
Therefore, the criterion is of the form,
\begin{eqnarray}
\label{ex3-(-1)-str}
{3 |\rho_{1, 64}|} \le \textrm{Tr}[W_{(-1)-\text{str}} \cdot \rho],
\end{eqnarray}
where $W_{(-1)-\text{str}}=\textrm{diag}\{a_1,\cdots\}$ is the diagonal matrix with $a_{2} = a_{3} = a_{4} = a_{5} = a_{6} = a_{7} = a_{9} = a_{10} = a_{11} = a_{13} = a_{16} = a_{17} = a_{18} = a_{19} = a_{21} = a_{24} = a_{25} = a_{28} = a_{30} = a_{31} = a_{32} = a_{33} = a_{34} = a_{35} = a_{37} = a_{40} = a_{41} = a_{44} = a_{46} = a_{47} = a_{48} = a_{49} = a_{52} = a_{54} = a_{55} = a_{56} = a_{58} = a_{60} = a_{59} = a_{61} = a_{62} = a_{63} = \frac{1}{2}$.

If $\rho$ is $3$-partitionable, $\Gamma_{3-\text{part}}^{\max}$ is a set of partitions of types $\{ 2, 2, 2 \}$,  $\{ 3, 2, 1 \}$ and $\{ 4, 1, 1 \}$. {For each partition in the set of partitions of type $\{ 2, 2, 2 \}$, we also consider 15 inequalities as before. For each partition in the set of partitions of type $\{ 4, 1, 1 \}$, we further consider six inequalities, each corresponding to a part that, as in the previous cases, contains one subsystem.  Employing the codes in \cite{Wu}, for each partition in the set of partitions of type $\{ 3, 2, 1 \}$, we sum the six inequalities twice.}
Therefore, we have
\begin{eqnarray}
\label{ex3-3part}
{3 |\rho_{1, 64}|} \le \textrm{Tr}[W_{3-\text{part}} \cdot \rho],
\end{eqnarray}
where $W_{3-\text{part}}=\textrm{diag}\{a_1,\cdots\}$ is the diagonal matrix with $a_{2} = a_{3} = a_{5} = a_{9} = a_{17} = a_{32} = a_{33} = a_{48} = a_{56} = a_{60} = a_{62} = a_{63} = 1$ and $a_{4} = a_{6} = a_{7} = a_{10} = a_{11} = a_{13} = a_{16} = a_{18} = a_{19} = a_{21} = a_{24} = a_{25} = a_{28} = a_{30} = a_{31} = a_{34} = a_{35} = a_{37} = a_{40} = a_{41} = a_{44} = a_{46} = a_{47} = a_{49} = a_{52} = a_{54} = a_{55} = a_{58} = a_{59} = a_{61} = \frac{1}{2}$.

If $\rho$ is $0$-stretchable, $\Gamma_{0-\text{str}}^{\max}$ is a set of partitions of types $\{ 2, 2, 2 \}$ and $\{ 3, 2, 1 \}$. The criteria is given by the inequality (\ref{ex3-3part}).

The inequalities (\ref{ex3-5part}, \ref{ex3-4part}, \ref{ex3-(-1)-str}, \ref{ex3-3part}) are all derived using the approach of fixing the right-hand side of the inequality and seeking a larger left-hand side.

If $\rho$ is $2$-producible, $\Gamma_{2-\text{prod}}^{\max}$ is a set of partitions of type $\{ 2, 2, 2 \}$.
{By the codes in \cite{Wu}, one of the criteria is}
\begin{eqnarray}
\label{ex3-2pro}
{|\rho_{1, 64}|} \le \textrm{Tr}[W_{2-\text{prod}} \cdot \rho],
\end{eqnarray}
where $W_{2-\text{prod}}=\textrm{diag}\{a_1,\cdots\}$ is the diagonal matrix with $a_{4} = a_{7} = a_{11} = a_{19} = a_{30} = a_{35} = a_{46} = a_{54} = a_{58} = a_{61} = \frac{1}{2}$.

If $\rho$ is $3$-producible, $\Gamma_{3-\text{prod}}^{\max}$ is a set of partitions of types $\{ 2, 2, 2 \}$ and $\{ 3, 3 \}$.
{By the codes in \cite{Wu}, one of the criteria is}
\begin{eqnarray}
\label{ex3-3pro}
{|\rho_{1, 64}| }\le \textrm{Tr}[W_{3-\text{prod}} \cdot \rho],
\end{eqnarray}
where $W_{3-\text{prod}}=\textrm{diag}\{a_1,\cdots\}$ is the diagonal matrix with $a_{4} = a_{7} = a_{8} = a_{11} = a_{12} = a_{14} = a_{15} = a_{19} = a_{20} = a_{22} = a_{23} = a_{26} = a_{27} = a_{29} = a_{30} = a_{35} = a_{36} = a_{38} = a_{39} = a_{42} = a_{43} = a_{45} = a_{46} = a_{50} = a_{51} = a_{53} = a_{54} = a_{57} = a_{58} = a_{61} = \frac{1}{2}$

If $\rho$ is $1$-stretchable, $\Gamma_{1-\text{str}}^{\max}$ is a set of partitions of types $\{ 2, 2, 2 \}$, $\{ 3, 3 \}$ and $\{ 4, 1, 1 \}$.{By the codes in \cite{Wu}, one of the criteria is}
\begin{eqnarray}
\label{ex3-1str}
{|\rho_{1, 64}|} \le \textrm{Tr}[W_{1-\text{str}} \cdot \rho],
\end{eqnarray}
where $W_{1-\text{str}}=\textrm{diag}\{a_1,\cdots\}$ is the diagonal matrix with $a_{2} = a_{3} = a_{4} = a_{5} = a_{7} = a_{8} = a_{9} = a_{11} = a_{12} = a_{14} = a_{15} = a_{17} = a_{19} = a_{20} = a_{22} = a_{23} = a_{26} = a_{27} = a_{29} = a_{30} = a_{35} = a_{36} = a_{38} = a_{39} = a_{42} = a_{43} = a_{45} = a_{46} = a_{48} = a_{50} = a_{51} = a_{53} = a_{54} = a_{56} = a_{57} = a_{58} = a_{60} = a_{61} = a_{62} = a_{63} = \frac{1}{2}$.

If $\rho$ is $4$-producible or $2$-stretchable, $\Gamma_{4-\text{prod}}^{\max}$ is a set of partitions of types $\{ 4, 2 \}$ and $\{ 3, 3 \}$, and the criterion is given by
\begin{eqnarray}
\label{ex3-4pro}
{|\rho_{1, 64}| }\le \textrm{Tr}[W_{4-\text{prod}} \cdot \rho],
\end{eqnarray}
where $W_{4-\text{prod}}^{P}=\textrm{diag}\{a_1,\cdots\}$ is the diagonal matrix with $a_{4} = a_{6} = a_{7} = a_{8} = a_{10} = a_{11} = a_{12} = a_{13} = a_{14} = a_{15} = a_{16} = a_{18} = a_{19} = a_{20} = a_{21} = a_{22} = a_{23} = a_{24} = a_{25} = a_{26} = a_{27} = a_{28} = a_{29} = a_{30} = a_{31} = a_{34} = a_{35} = a_{36} = a_{37} = a_{38} = a_{39} = a_{40} = a_{41} = a_{42} = a_{43} = a_{44} = a_{45} = a_{46} = a_{47} = a_{49} = a_{50} = a_{51} = a_{52} = a_{53} = a_{54} = a_{55} = a_{57} = a_{58} = a_{59} = a_{61} = \frac{1}{2}$.

If $\rho$ is $5$-producible or $3$-stretchable, $\Gamma_{4-\text{prod}}^{\max}$ is a set of partitions of types $\{ 5, 1 \}$, $\{ 4, 2 \}$ and $\{ 3, 3 \}$, and the criterion is
\begin{eqnarray}
\label{ex3-5pro}
{|\rho_{1, 64}|} \le \textrm{Tr}[W_{5-\text{prod}} \cdot \rho],
\end{eqnarray}
where $W_{5-\text{prod}}=\textrm{diag}\{a_1,\cdots\}$ is the diagonal matrix with $a_{2} = a_{3} = a_{4} = a_{5} = a_{6} = a_{7} = a_{8} = a_{9} = a_{10} = a_{11} = a_{12} = a_{13} = a_{14} = a_{15} = a_{16} = a_{17} = a_{18} = a_{19} = a_{20} = a_{21} = a_{22} = a_{23} = a_{24} = a_{25} = a_{26} = a_{27} = a_{28} = a_{29} = a_{30} = a_{31} = a_{32} = a_{33} = a_{34} = a_{35} = a_{36} = a_{37} = a_{38} = a_{39} = a_{40} = a_{41} = a_{42} = a_{43} = a_{44} = a_{45} = a_{46} = a_{47} = a_{48} = a_{49} = a_{50} = a_{51} = a_{52} = a_{53} = a_{54} = a_{55} = a_{56} = a_{57} = a_{58} = a_{59} = a_{60} = a_{61} = a_{62} = a_{63} = \frac{1}{2}$.

The observables $W_P$ and $W_{k-\text{prod}},$ $W_{k-\text{part}}$ and $W_{k-\text{str}}$ are expressed as follows
\begin{align}
\begin{aligned}
&W_P = \bigotimes_{i=1}^n H_i^{+} + \bigotimes_{i=1}^n H_i^{-}\\
&W_{1-\text{prod}} = \frac{1}{2}(Z_1^{+} Z_2^{+} Z_3^{+} Z_4^{+} Z_5^{+} Z_6^{-} + Z_1^{-} Z_2^{-} Z_3^{-} Z_4^{-} Z_5^{-} Z_6^{+})\\
&W_{2-\text{prod}}=\frac{1}{2}(\Delta_1+X\Delta_1X)\\
&W_{3-\text{prod}}=\frac{1}{2}(\Delta_1+X\Delta_1X+\Delta_2+X\Delta_2X+\Delta_3+X\Delta_3X)\\
&W_{4-\text{prod}} = \frac{\mathbb{I}_{64}}{2} - \frac{1}{2} \bigotimes_{k=1}^6 Z_k^{+} -\frac{1}{2} \bigotimes_{k=1}^6 Z_k^{-} - \\
&\quad\quad\frac{1}{2} \sum_{k=1}^6 Z_k^{-} \otimes (\bigotimes_{j \neq k} Z_{j}^{+} ) - \frac{1}{2} \sum_{k=1}^6 Z_k^{+} \otimes (\bigotimes_{j \neq k} Z_j^{-})
\end{aligned}
\end{align}

\begin{align*}
\begin{aligned}
&W_{5-\text{prod}} = \frac{\mathbb{I}_{64}}{2} - \frac{1}{2} \bigotimes_{k=1}^6 Z_k^{+} -\frac{1}{2} \bigotimes_{k=1}^6 Z_k^{-}\\
&W_{1-\text{str}}=\frac{\mathbb{I}_{64}}{2}-\frac{1}{2}(\Delta_4+X\Delta_4X+\otimes_{i=2}^6Z_i^++\otimes_{i=2}^6Z_i^-)
\end{aligned}
\end{align*}

\begin{align*}
\begin{aligned}
&W_{-1-\text{str}}=\frac{\mathbb{I}_{64}}{2}-\frac{1}{2}(\Delta_2+X\Delta_2X+\Delta_3+X\Delta_3X+\\
&\quad\quad\otimes_{i=1}^6Z_i^++\otimes_{i=1}^6Z_i^-)\\
&W_{3-\text{part}}=\frac{\mathbb{I}_{64}}{2}+\frac{1}{2}(\Delta_5+X\Delta_5X-\Delta_2-X\Delta_2X-\Delta_3\\
&\quad\quad-X\Delta_3X-\otimes_{i=1}^6Z_i^+-\otimes_{i=1}^6Z_i^-)\\
\end{aligned}
\end{align*}
\begin{align}
\begin{aligned}
&W_{4-\text{part}}=W_{5-\text{part}}^{P}=\sum_{k=1}^6 Z_k^{-} \otimes (\bigotimes\limits_{j=1, j \neq k}^6 Z_j^{+})\\
&\quad\quad+ \sum_{k=1}^6 Z_k^{+} \otimes (\bigotimes\limits_{j=1, j \neq k}^6 Z_j^{-})
\end{aligned}
\end{align}
with
\begin{align}
\begin{aligned}
&X=\otimes_{i=1}^6\sigma_x^{(i)}\\
& \Delta_1 = Z_1^{+} Z_2^{+} (Z_3^{+} Z_4^{+} Z_5^{-} Z_6^{-} + Z_3^{+} Z_4^{-} Z_5^{-} Z_6^{+} + Z_3^{-} Z_4^{+} Z_5^{-} Z_6^{+})\\
&\quad\quad+ (Z_1^{+} Z_2^{-} + Z_1^{-} Z_2^{+}) Z_3^{+} Z_4^{+} Z_5^{-} Z_6^{+} \\
&\Delta_2 = Z_1^{+} Z_2^{+} (Z_3^{+} Z_4^{+} Z_5^{+} Z_6^{-} + Z_3^{+} Z_4^{+} Z_5^{-} Z_6^{+} + Z_3^{+} Z_4^{-} Z_5^{+} Z_6^{+}\\
&\quad\quad+ Z_3^{-} Z_4^{+} Z_5^{+} Z_6^{+})\\
&\Delta_3= Z_1^{+} Z_2^{-} (Z_3^{+} Z_4^{+} Z_5^{-} Z_6^{-} + Z_3^{+} Z_4^{-} Z_5^{-} Z_6^{+} + Z_3^{-} Z_4^{+} Z_5^{-} Z_6^{+}\\
&\quad\quad+ Z_3^{-} Z_4^{-} Z_5^{+} Z_6^{+}+ Z_3^{-} Z_4^{+} Z_5^{+} Z_6^{-}+ Z_3^{+} Z_4^{-} Z_5^{+} Z_6^{-})\\
&\Delta_4=(Z_1^+Z_2^-+Z_1^-Z_2^+)(Z_3^{+} Z_4^{+} Z_5^{+} Z_6^{-}+Z_3^{+} Z_4^{-} Z_5^{+} Z_6^{+}\\
&\quad\quad+Z_3^{-} Z_4^{+} Z_5^{+} Z_6^{+})+Z_1^+Z_2^+(Z_3^+Z_4^-Z_5^+Z_6^-\\
&\quad\quad+Z_3^-Z_4^+Z_5^+Z_6^-+Z_3^-Z_4^-Z_5^+Z_6^+Z_3^-Z_4^-Z_5^-Z_6^-)\\
&\Delta_5=\sum\limits_{i=1}^6 Z_i^-\otimes(\bigotimes\limits_{j=1,j\neq i}^6 Z_j^+)
\end{aligned}
\end{align}
and $\sigma_x^{(i)}$ being the pauli matrice in the $i$-th subsystem.

\bibliography{biblio}
\end{document}